\documentclass[traditabstract,referee]{aa}
\usepackage{txfonts}
\usepackage{natbib}
\usepackage{graphicx}
\begin{document}

\newcommand{\lele}[3]{{#1}\,$\le$\,{#2}\,$\le$\,{#3}}
\newcommand{\tauav}[1]{$\tau_{#1}$/A$_{\rm V}$}
\newcommand{\dr}{2012\,DR$_{30}$}
\newcommand{\redchi}{$\chi^2_r$}
\newcommand{\ti}{$\rm J\,m^{-2}\,s^{-0.5}\,K^{-1}$}
\sloppy

\title{A portrait of the extreme Solar System object 
\dr\thanks{{\it Herschel} is an ESA space observatory 
with science instruments provided by
European-led Principal Investigator consortia and with important participation from NASA.}
}
%
%
\author{ Cs.~Kiss\inst{1} 
\and Gy.~Szab\'o\inst{1,2,3} 
\and J.~Horner\inst{4}
\and B.C.~Conn\inst{5}
\and T.G.~M\"uller\inst{6}
\and E.~Vilenius\inst{6}
\and K.~S\'arneczky\inst{1,2}
\and L.L.~Kiss\inst{1,2,7}
\and M.~Bannister\inst{8}
\and D.~Bayliss\inst{8}
\and A.~P\'al\inst{1,9}
\and S.~G\'obi\inst{1}
\and E.~Vereb\'elyi\inst{1}
\and E.~Lellouch\inst{10}
\and P.~Santos-Sanz\inst{11}
\and J.L.~Ortiz\inst{11}
\and R.~Duffard\inst{11}
\and N.~Morales\inst{11}
}

\institute{ Konkoly Observatory, MTA CSFK, H-1121 Budapest, Konkoly Th.M. \'ut 15-17., Hungary
    \and
    ELTE Gothard-Lendület Research Group, H-9700 Szombathely, Szent Imre herceg \'ut 112, Hungary
    \and
    Dept. of Exp. Physics \&{} Astronomical Observatory, University of Szeged, H-6720 Szeged,
    Hungary
    \and
    Department of Astrophysics, School of Physics, University of New South Wales,
    Sydney, NSW 2052, Australia
    \and
    Max-Planck-Institut f\"ur Astronomie, K\"onigstuhl 17, 69117 Heidelberg, Germany 
    \and
    Max-Planck-Institut f\"ur extraterrestrische Physik, Giessenbachstrasse,
    85748 Garching, Germany
    \and
    Sydney Institute for Astronomy, School of Physics, A28, The University of Sydney, 
    NSW 2006, Australia
    \and
    Research School of Astronomy and Astrophysics, 
    the Australian National University, ACT 2612, Australia
	\and
    Department of Astronomy, Lor\'and E\"otv\"os University, 
    P\'azm\'any P\'eter s\'et\'any 1/A, H-1119 Budapest, Hungary
    \and
    Observatoire de Paris, Laboratoire d'\'Etudes Spatiales et
    d'Instrumentation en Astrophysique (LESIA),
    5 Place Jules Janssen, 92195 Meudon Cedex, France
    \and
    Instituto de Astrof\'\i{}sica de Andaluc\'\i{}a (IAA-CSIC)
    Glorieta de la Astronom\'\i{}a, s/n 18008 Granada, Spain
    }


\date{Received \today / Accepted \today}
\abstract{ \dr{} is a recently discovered Solar System object on a unique
orbit, with a high eccentricity of 0.9867, a perihelion distance 
of 14.54\,AU and a semi-major axis of 1109\,AU, in this respect
outscoring the vast majority of trans-Neptunian objects. 
We performed Herschel/PACS and optical photometry to uncover the size
{ and albedo of \dr{}, together with its} thermal and surface properties. 
The body is 185\,km in diameter and has a 
relatively low V-band geometric albedo of $\sim$8\%. Although the colours of the object
indicate that \dr{} is an $RI$ taxonomy class TNO or Centaur, 
we detected an absorption feature in the $Z$-band that is uncommon among
these bodies. A dynamical analysis of the target's orbit shows that \dr{} moves on a 
relatively unstable orbit and was most likely only recently placed on its 
current orbit from the most distant and still highly unexplored regions
of the Solar System. If categorised on dynamical grounds 
\dr{} is the largest Damocloid and/or high inclination Centaur 
observed so far.}

\maketitle

\section{Introduction \label{sect:introduction}}

\dr{} was discovered on February 26, 2012 (MPEC~2012-D67) as part of the
Siding Spring Survey\footnote{http://www.mso.anu.edu.au/~rmn/index.htm}.
Shortly after the discovery of \dr{}, it was realised that the object was 
identical to the object 2009\,FW$_{54}$. As a result of that identification, it has 
been possible for the object's orbit to be determined with relatively high precision, 
based on 142 observations made between February 2008 and April 2012.
The orbit is rather peculiar with a semi-major axis of a\,=\,1109\,AU, eccentricity of
e\,=\,0.9869 and inclination of i\,=\,78$\fdg$00. 
The object is currently close to its perihelion, q\,=\,14.54\,AU on 
its $\sim$37 thousand-year long orbit
(see also the Minor Planet Center page of 
\dr{}\footnote{http://www.minorplanetcenter.net/db\_search/ show\_object?object\_id=2012+dr30}). 



%
%
%

Following Gladman et al. (2008), \dr{} would most likely be classified 
as a Scattered Disk
Object. However, the Gladman et al. scheme was mainly focussed on understanding 
the behaviour of objects originating in the trans-Neptunian region, 
and so such a classification would 
naturally lead the reader to infer an origin for \dr{} in the trans-Neptunian 
population. On the other hand, Brasser et al. (2012) consider objects on orbits like 
that of \dr{} to be high inclination Centaurs, and, along with Emel'yanenko et al.
 (2005), have suggested that the origin for these objects could well be the inner Oort 
cloud, rather than the trans-Neptunian belt. Currently, only three objects 
are categorised as high inclination Centaurs: 2002\,XU$_{93}$, 
2008\,KV$_{42}$ and 2010\,WG$_9$ \citep{Brasser2012}.  

Considering the high eccentricity and inclination of \dr{}'s orbit,
another way to categorise it might be to consider it a high-perihelion, 
long-period comet since it shares many characteristics with those objects
(as its perihelion distance is $\sim$15\,AU, it is not classified as
having cometary dynamics according to the \citet{Gladman2008} 
scheme that requires T$_J$\,$<$\,3.05 and q\,$<$\,7.35\,AU).
Many objects moving on typical 
Centaur orbits were classified as comets on the basis of cometary activity at
discovery -- this has not, to date, been observed in the case of \dr{}.

{
Given the similarity of the orbit of \dr{} to those of the long period comets 
it is worth considering a different 
mechanisms proposed to emplace objects in such orbits. 
An object with an aphelion 
distance of $\sim$\,1000\,AU is typically considered to be too tightly bound to the Sun 
for its orbit to be significantly perturbed by the influence of the galactic tide or 
close encounters with passing stars (processes which are considered far more important 
at aphelion distances of $\sim$10,000\,AU, or greater). 
The orbital evolution of 
such objects is thought to currently be driven by chance encounters with the giant 
planets:
cometary bodies moving on short-period orbits within the outer Solar System are 
regularly scattered to longer period orbits -- or even entirely ejected from the Solar 
system (Horner et al., 2004b). 
Typically, however, objects on highly eccentric orbits (with aphelia at thousands, or 
tens of thousands of AU) are thought to have been be recently injected 
from the Oort cloud. The classical route through which such objects are 
emplaced to their current orbits involves 
their injection from the outer Oort cloud, where their orbits can be strongly modified 
by the influence of the galactic tide and the gravitational influence of passing stars 
(e.g. Wiegert \& Tremaine, 1999; Dones et al., 2004; Rickman et al., 2008). 

Whilst this mechanism is good at explaining the observed distribution of long-period 
comets with perihelion distances in the range 5--10\,AU, it has great difficulties 
explaining the existence of the high-inclination Centaurs (as described by e.g. 
Brasser et al., 2012). These objects display high perihelion distances (e.g. 14.54\,AU, 
in the case of \dr), which would typically be considered as too distant for Jupiter 
and Saturn to easily decouple the object's orbital evolution from the influence of 
passing stars and the galactic tide. The aphelion distance of these objects, however 
is significantly too small for them to have been 
injected to their current orbit by a passing star or by the galactic tide.
For this reason, a number of 
authors have proposed that such objects are instead sampling the inner Oort cloud 
population (e.g. Emel'yanenko et al., 2005; Brasser et al., 2012).

Since \dr{} falls very close to the tenuous q\,$\approx$\,15\,AU boundary between those 
objects that could theoretically be decoupled from the outer Oort cloud by the 
influence of Jupiter and Saturn, and those objects that could not be captured 
in this way, it is clearly had to definitively argue for one particular origin 
over the other. In either case, however, it seems reasonable that it could 
well be a relatively recent entrant to the inner reaches of the Solar System.
}

On dynamical grounds \dr{} also shows similarities to the group of Damocloids 
which are thought to be inactive Halley-type or long-period comets.
According to the definition given by \citet{Jewitt2005} objects in this
group have a Tisserand-parameter relative to Jupiter T$_J$\,$\le$\,$2$, and
indeed, this parameter for \dr{} is T$_J$\,=\,0.198. On the other hand, 
these objects have perihelion distances typically q\,$\lesssim$\,5\,AU and
are small (H$_V$\,$>$\,10) in most cases, unlike \dr{} (H$_V$\,$\approx$\,7). 
Some of the few exceptions are the three 
high incination Centaurs mentioned above which 
are also Damocloids according to their T$_J$-s. These objects should be
relatively large (H$_V$\,$\approx$\,8\fm0-9\fm0). A reliable size estimate 
is only available for 2002\,XU$_{93}$, which has 
an effective diameter of 164$\pm$9\,km, based
on thermal emission measurements with the Herschel Space Observatory
\citep{SS2012}. The surfaces of the Damocloids are among the darkest 
ones known in the Solar System --
the objects for which albedos are known so far all have p$_V$\,$\approx$\,0.04
\citep[see][]{Jewitt2005,SS2012}.

In the case of \dr{} it is clearly interesting to consider whether 
there might be any observational evidence that could support one possible
origin over another, particularly if \dr{}
could be recently placed to its current orbit from the inner Oort cloud.
Apart from dynamical behaviour, evidence may also come e.g. from 
surface characteristics reflecting the different evolutionary paths this
object might have taken. In this paper we investigate the basic 
physical properties of \dr{} with the help of thermal emission and 
optical follow-up
photometry observations and try to relate these to the dynamics of the
orbit of this peculiar object. 

\begin{table*}[Ht!]
\caption{Summary of Herschel/PACS, WISE and MPG/ESO 2.2m observations. 
The columns of the table are:
(1) Telescope and instrument; 
(2) Observation identifier; 
(3) Photometric band(s), identified with the nominal wavelength(s);
(4) Start date and time of the observation (UTC);
(5) Start time of the observation (Julian Date);
(6) Heliocentric distance (AU);
(7) Target distance (AU); 
(8) Phase angle (deg). 
The data of the last three columns were taken from NASA's Horizons Database. 
Herschel observations lasted for 1414\,s. }
\label{table:herschelobs}
\begin{tabular}{cccccccc}
\hline
Telescope      & OBSID    &  Band   & Start-time  & Start time    & r    & $\Delta$ & $\alpha$ \\
 \& instrument &          &         &              &(JD-2450000) & (AU) & (AU)     &  (deg)   \\
\hline
 Herschel/PACS & 13421246148 &   70/160$\mu$m & 2012-May-25 23:31 &  6073.479 &  14.671 & 14.509 & 3.971 \\
 Visit-1       & 13421246149 &   70/160$\mu$m & 2012-May-25 23:55 &  6073.496 &  14.671 & 14.509 & 3.971 \\
               & 13421246150 &  100/160$\mu$m & 2012-May-26 00:20 &  6073.513 &  14.671 & 14.509 & 3.972 \\
               & 13421246151 &  100/160$\mu$m & 2012-May-26 00:45 &  6073.531 &  14.671 & 14.509 & 3.972 \\ \hline
 Herschel/PACS & 13421246215 &   70/160$\mu$m & 2012-May-27 21:48 &  6075.408 &  14.672 & 14.539 & 3.985 \\
 Visit-2       & 13421246216 &   70/160$\mu$m & 2012-May-27 22:13 &  6075.425 &  14.672 & 14.539 & 3.985 \\
               & 13421246217 &  100/160$\mu$m & 2012-May-27 22:38 &  6075.443 &  14.672 & 14.540 & 3.985 \\
               & 13421246218 &  100/160$\mu$m & 2012-May-27 23:02 &  6075.459 &  14.672 & 14.540 & 3.986 \\ 
\hline
 WISE     &  --   &  11/22$\mu$m &  2010-May-25 20:37  &  5342.359  &  14.600 & 14.482 & 3.964 \\
 W3/W4    &  --   &  11/22$\mu$m &  2010-May-26 04:33  &  5342.690  &  14.600 & 14.488 & 3.966 \\
          &  --   &  11/22$\mu$m &  2010-May-26 06:09  &  5342.756  &  14.600 & 14.489 & 3.966 \\
          &  --   &  11/22$\mu$m &  2010-May-26 07:44  &  5342.822  &  14.600 & 14.490 & 3.966 \\
          &  --   &  11/22$\mu$m &  2010-May-26 09:19  &  5342.888  &  14.600 & 14.491 & 3.967 \\
          &  --   &  11/22$\mu$m &  2010-May-26 10:54  &  5342.954  &  14.600 & 14.492 & 3.967 \\
          &  --   &  11/22$\mu$m &  2010-May-26 14:05  &  5343.087  &  14.600 & 14.494 & 3.968 \\                               
\hline
 2.2m/WFI      &    --         &   B (451\,nm) & 2012-Jun-07 00:51 & 6085.036 &  14.671 & 14.520 & 3.975 \\
 2.2m/WFI      &    --         &   V (540\,nm) & 2012-Jun-06 23:55 & 6084.996 &  14.671 & 14.520 & 3.975 \\
 2.2m/WFI      &    --         &   R (652\,nm) & 2012-Jun-07 01:00 & 6085.042 &  14.671 & 14.520 & 3.975 \\
 2.2m/WFI      &    --         &   I (784\,nm) & 2012-Jun-07 01:10 & 6085.049 &  14.671 & 14.520 & 3.975 \\
 2.2m/WFI      &    --         &   Z (964\,nm) & 2012-Jun-07 01:29 & 6085.062 &  14.671 & 14.520 & 3.975 \\                 
\hline 
\end{tabular}
\end{table*}

\section{Observations and data reduction}

\subsection{Herschel/PACS observations}

Thermal emission of \dr{} was observed
with the PACS photometer camera 
\citep{PACS} of the Herschel Space Observatory \citep{Herschel}
using the time awarded in a DDT proposal exclusively for \dr{} 
(proposal ID: DDT\_ckiss\_2). The observations were performed in
mini-scanmap mode, homogeneously covering
a field of roughly 1\arcmin\, in diameter (Fig. 1). This mode 
is suited for our needs and offers more sensitivity than
other observation modes of the PACS photometer \citep{MuellerSDP}. 

The reduction of raw data was performed in the Herschel Interactive
Pipeline Environment (HIPE, Ott 2011) using an optimized version of the PACS 
bright point source pipeline script without the application of 
proper motion correction due to slow motion of the target 
relative to the telescope beam size during a single OBSID ($<$1\arcsec). 
We derived single epoch co-added images in each Herschel/PACS band, as well as 
differential and "double-differential" images combining the data of the 
two epochs in order to get rid of the confusion due to the sky background. 
Differential images are created by subtracting the co-added image of the
second visit from that of the first visit image of the same band. 
This eliminates the background and leaves a positive and a 
negative beam of the target on the image. The ideal matching of the two image 
frames is obtained using a fluctuation minimalisation method.  
To create a double differential image, a copy of the differential
image is folded and shifted in a way that the positive beams of the
two visits are matched in position and then co-added, providing a positive beam 
with the average flux of the target, plus two negative beams on the sides 
with fluxes about half the central, positive beam (the photometry 
is performed on the central beam). 
The procedure to create these images, and also the photometry of the target  
was performed in the same way as it is described in detail in 
\citet{Pal2012,Vilenius2012,SS2012}. The photometric fluxes we obtained 
are summarized in Table~\ref{table:herschelres}. 


\begin{figure*}[ht!!!!!!]
\hbox{ \includegraphics[width=5.5cm]{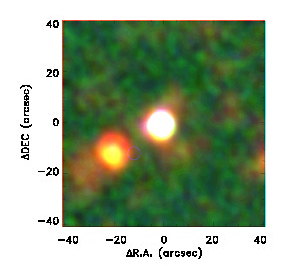}
	   \includegraphics[width=5.5cm]{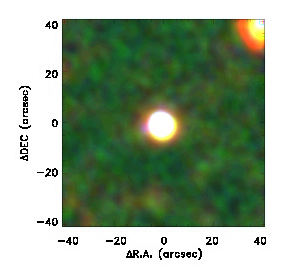}
	   \includegraphics[width=5.5cm]{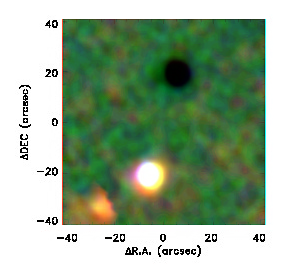}
}
\caption[]{False colour images of
Herschel/PACS photometer with the nominal wavelengths of 
70 (blue), 100 (green) and 160\,$\mu$m (red). Left and middle panels 
represent the Visit-1 and Visit-2 images, respectively, in the corresponding
bands, while the right column shows the differential image.
\dr{} is in the centre of the Visit-1 and Visit-2 images, and is seen
as a pair of bright/dark spots on the differential image.}
\end{figure*}
\begin{table}[Ht!]
\caption[]{Summary table of the Herschel photometry results. Visit-1 and Visit-2 
refers to the two epochs of observations (see Table~\ref{table:herschelobs}). 
Visit-1 and Visit-2 differential fluxes are derived from the 
\emph{differential}, background eliminated maps, performing photometry on
the positive and negative beams of the target independently. 
The \emph{double differential} fluxes are derived from the double 
differential maps performing photometry on the central, combined beam.}
\label{table:herschelres}
\begin{tabular}{ccccc}
\hline
Band  & Epoch  & F$_{\rm coadd}$ & F$_{\rm diff}$ & F$_{\rm ddiff}$\\ 
($\mu$m)  &   &  (mJy)         &  (mJy)         & (mJy) \\ \hline
70  & Visit-1 & 92.83$\pm$1.45 & 91.80$\pm$0.88 & 88.56$\pm$0.62 \\
70  & Visit-2 & 89.41$\pm$1.25 & 87.18$\pm$0.88 &                \\ \hline
100 & Visit-1 & 64.12$\pm$1.46 & 63.87$\pm$1.20 & 62.40$\pm$0.91 \\
100 & Visit-2 & 66.22$\pm$1.94 & 62.08$\pm$1.20 &                \\ \hline
160 & Visit-1 & 37.83$\pm$2.38 & 32.82$\pm$1.34 & 34.48$\pm$1.13 \\
160 & Visit-2 & 40.37$\pm$2.66 & 36.52$\pm$1.34 &                \\ \hline
\end{tabular}
\end{table}



\subsection{MPG/ESO 2.2\,m optical follow-up at La Silla}

We performed photometric measurements of \dr{} in a service mode 
Max-Planck-Institut f\"ur Astronomie (MPIA) DDT 
observation at La Silla Observatory, Chile, using the MPG/ESO 2.2m telescope 
(PID 089.A--9031(A)). 

Data were taken with the Wide Field Imager on June 6, 2012 and consist of images
in filters B (BB\#B/123\_ESO878, 60s), V (BB\#V/89\_ESO843, 60s), R
(BB\#Rc/162\_ESO844, 60s), I (BB\#I/203\_ESO879, 60s) and Z
(BB\#Z+/61\_ESO846, 280s). We used three individual exposures for the
B, V, R and I filters each, while the Z had six exposures. The
Cambridge Astronomical Survey Unit pipeline
\citep{2001NewAR..45..105I} was used to combine
the bias frames and dome flat fields, obtained on the same night, into
master bias and master flat frames which were then used for the bias
subtraction and flat fielding. Preliminary source extraction and
astrometry, crossmatched with the 2MASS point source catalogue
\citep{2006AJ....131.1163S}, were determined for
all the reduced frames allowing the multiple observations per filter
to be each stacked into a deeper image. The multiple I and Z frames
were median combined to form fringe frames which were then used to
defringe the individual I and Z frames before stacking. The source
extraction and astrometry were then repeated and refined on the
stacked frames with an accuracy of better than 0$\farcs 2$. The
seeing was typically better than 1$\farcs 3$ during the observations.

$BVRI$ standard stars were taken from SA~104 in the Landolt (1992)
catalog to measure the photometric nature of the night. A field
overlapping with SDSS was observed to provide a reference for the Z band.
$Z$ images were then transformed to SDSS $z$ magnitudes following two
methods: (i) synthetic $z$ magnitudes calculated for field stars
\citep{2006AJ....132..989R}, and (ii) standard
stars in the SDSS stripe \#{}1540 (RA= 11:29:30  Dec= --07:00:02) in
15$^\circ$ vicinity of \dr. The two methods gave consistent
zeropoints within $\pm 0.02$~mag. Atmospheric extinction coefficients
were taken from ESO website for La Silla\footnote{www.ls.eso.org}.


\subsection{WISE observations of \dr{}}

\dr{} was not seen by the Wide-field Infrared Survey Explorer \citep[WISE][]{Wise1}
according to the MPC entries and the WISE search tools.
It might be that \dr{} was not yet known at the time the Solar System search
programmes were executed \citep[see e.g. ][]{Wise2} or that it simply was not
recognised as moving target due to its slow apparent sky motion
of only 1-5$^{\prime \prime}$/h.
But based on the PACS measurements and flux extrapolations to
the WISE W3 (11.56\,$\mu$m) and W4 (22.09\,$\mu$m) bands it became
clear that WISE must have seen \dr{}.

We found the source {\it J103104.77+005635.9} within 1$^{\prime \prime}$
of the \dr{} path in the WISE all-sky source 
catalog\footnote{http://irsa.ipac.caltech.edu/applications/wise/}.
The W3 and W4 magnitudes are 12.037 and 6.900, respectively. The
WISE image catalog shows a sequence of several detections
in W4, but the source is clearly moving and appears elongated.
We therefore used the WISE all-sky single exposure (L1b) source
table which includes several source detections along the apparent
sky path of \dr{}. The seven W3-detections (with S/N $>$2) and
the nine W4-detections were taken in the period
MJD 55341.85947 -- 55342.58708 (mean: 55342.24166). The weighted
mean W3-band magnitude was 10\fm88$\pm$0\fm22 and the typical
S/N was about 3, the weighted mean W4-band magnitude was
5\fm95$\pm$0\fm16 and the typical S/N was about 10.

We converted the observed magnitudes via the Vega model spectrum
into fluxes. Due to the red colour of \dr{} (compared to
the blue calibration stars) there is an additional correction needed
\citep[see][]{Wise1} and the W3-flux has to be increased by 17\%
and the W4-flux has to be lowered by 9\%. It is also necessary to
apply a colour correction, which we calculated via a TPM prediction
of the spectral energy distribution of \dr{} (corresponding
roughly to a black body temperature of slightly above 100\,K). The colour
correction factors are 2.35 ($\pm$ 10\%) in W3 and 1.00 ($\pm$ 1\%)
in W4. The large error for the W3 colour correction is due to the
uncertain shape of the SED at these short wavelengths. We also added
a 10\% error for the absolute flux calibration in W3 and W4 which was
estimated from the discrepancy between red and blue calibrators
\citep{Wise1} and we combined all errors quadratically.
The final mono-chromatic flux densities at the WISE reference
wavelengths are listed in Table~\ref{table:inputfluxes}.




\begin{table*}
\caption[]{Summary on the calculation of input fluxes and flux uncertainties. The columns are: 
(1) Instrument; (2) Band and/or reference wavelength; (3) Colour correction factor; 
(4) flux correction; (5) absolute calibration uncertainty; (6) measured flux;
(7) monochromatic flux used in the thermal models 
(see also Sect.~\ref{sect:fluxunc}. Note that in the case 
of the Herschel/PACS the uncertainties of the input fluxes are dominated
by the absolute calibration uncertainties.
\label{table:inputfluxes}}
\begin{tabular}{ccccccc}
\hline
Instrument 		&  Band/Wavelength  &  C$_{\lambda}$  & r$_{corr}$ & r$_{cal}$ & F$_{meas}$ & F$_{inp}$\\ 
			&  ($\mu$m)               &                             &                  &                 &   (mJy)      &   (mJy)    \\
\hline      
WISE (W3)		& 11.6\,$\mu$m &   2.35$\pm$0.24   &  1.17    &  0.10   &   1.29$\pm$0.26   &  0.64$\pm$0.16  \\
WISE (W4)		& 22.1\,$\mu$m &   1.00$\pm$0.01   &  0.91    &  0.10   &  34.50$\pm$5.10   &  31.4$\pm$5.6    \\
Herschel/PACS	&  70\,$\mu$m  &   0.99$\pm$0.02   &  1.00    &  0.05   &  88.56$\pm$0.62   &  89.45$\pm$3.29 \\
Herschel/PACS	& 100\,$\mu$m  &   1.01$\pm$0.02   &  1.00    &  0.05   &  62.40$\pm$0.91   &  61.78$\pm$2.39 \\
Herschel/PACS	& 160\,$\mu$m  &   1.045$\pm$0.020  &  1.00   &  0.05   &  34.48$\pm$1.13   &  33.00$\pm$2.07 \\
\hline
\end{tabular}
\end{table*}

\subsection{Light curve observations \label{sect:lightcurve}}

We obtained visible-light imagery of 2012 DR$_{30}$ with queue observations 
on the 2\,m Faulkes South telescope at Siding Spring Observatory, NSW. 
The Spectral camera of Faulkes South that we used has a plate scale of 
0.304 arcsec/pixel, with a 4k$\times$4k CCD array, giving a field of view of 
10.5\,arcmin. 

We observed 2012 DR$_{30}$ through the Sloan Digital Sky Survey 
r' filter on four nights, as detailed in Table~\ref{table:dr30astro}. 
On each night, consecutive 300 s exposures were obtained with 2012 DR$_{30}$ 
centred on the array; the 2x2 binning set a readout time between exposures of 22\,s. 
The rate of motion of 2012 DR$_{30}$ across the sky was 
0.9\arcsec/hr (a quarter-pixel, 75 milliarcsec, in 5 minutes), 
which kept it well within the seeing disk of $\sim2$\arcsec over each integration. 
Due to the low altitude of 2012 DR$_{30}$, the longest continuous set of observations 
were made over less than two hours. 34 integrations were made in total, excluding 
nine images where the seeing deteriorated below where 2012 DR$_{30}$ could be detected.

Bias subtraction, flatfielding and astrometry were provided by the standard Faulkes 
queue pipeline. The flatfielding varied in quality due to proximity to the Moon on 
some of the nights, creating a pocked "golf course" effect in places, but 2012 DR$_{30}$ 
did not fall on any problematic locations in these observations.
We then used SExtractor \citep{1996A&AS..117..393B} to obtain the flux of 2012 DR$_{30}$ 
from the reduced images. 

The 10 arcmin field of view provided ample suitably bright field stars, from which we 
selected eleven to act as comparison stars and cancel the effects of atmospheric 
variability (Table~\ref{table:compstars}). The selection was based on their photometric 
stability, lack of saturation, no blends or other close stars, and their spatial 
distribution on the field; the mutual relative photometry of these eleven stars varied 
by less than 0.01 mag across all the observations. We then measured the differential 
brightness variation of 2012\,DR$_{30}$ against these stars. 

The field on which 2012\,DR$_{30}$ fell during our observations 
(R.A. $= 10^{h}16^{m}$, decl. $= -17^{\circ}05^{\prime}$) was too far south to be within 
the SDSS photometric catalog. We instead used the AAVSO Photometric All-Sky Survey 
(APASS) Data Release 6 survey catalog for absolute calibration of the magnitudes of 
the comparison stars. Four had matches within 2.2\arcsec-0.5\arcsec in the catalogue 
(Table~\ref{table:compstars}); these matches were confirmed by visual inspection 
of the images. 
The known catalogue magnitudes allowed us to tie the zeropoint of the differential 
variation of 2012 DR$_{30}$ to an absolute magnitude. The scatter in the shift between 
the observed stellar magnitude and the catalogue value for the four comparison stars 
was 0.2 magnitudes; we therefore note that the internal precision in the relative 
photometry is much greater and provides a better measure of the variability 
of 2012\,DR$_{30}$.

\begin{figure*}
\includegraphics[width=\textwidth]{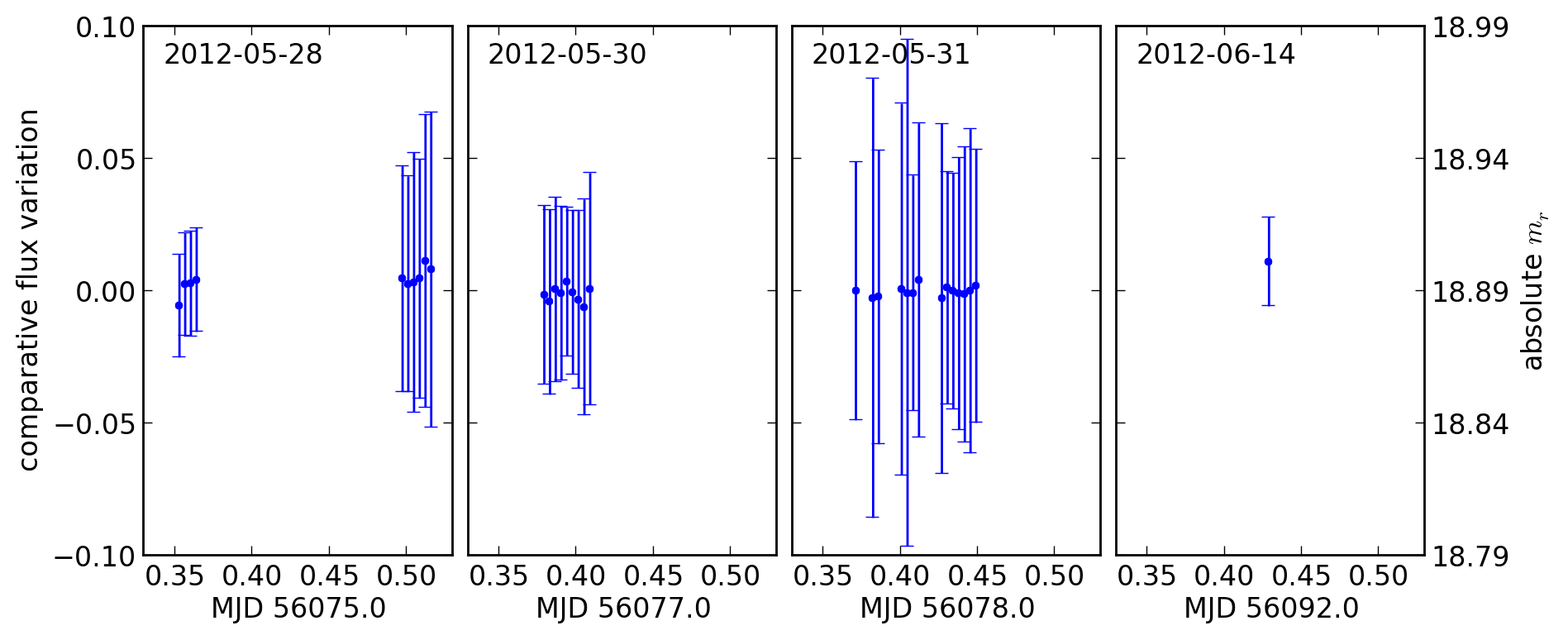}
\caption{Comparative photometry of  2012 DR$_{30}$ over four nights. 
The error bars are given relative to the internally consistent photometry 
(left ordinate), and is dominated by the photon noise, found via 
SExtractor's FLUXERR\_APER (RMS error vector for aperture flux) values. 
The absolute magnitude (right ordinate) is tied to the APASS catalogue, 
which had up to an 0.2 $m_{r}$ variation in the shift required for the 
standard stars, so it is provided for guidance rather than high precision.}
\label{fig:dr30lightcurve}
\end{figure*}

We used these comparative photometric measurements of 2012\,DR$_{30}$ to construct 
a light curve (Fig.~\ref{fig:dr30lightcurve}). This showed very little variation
and indicated an upper limit of 0\fm004 1-sigma variability when only the
standard deviation of the target's $r$ band brightness values are considered.
Note that the uncertainty of the individual measurements are dominated 
by the error of absolute calibration with a typical value 0\fm05
(see also Fig.~\ref{fig:dr30lightcurve}).  
We tried to fit a rotation period but the periodogram showed aliases 
only at one- and half-day periods, which would be spurious effects from the cadence 
of the observations. As a summary of these observations we can conclude 
that we have not been able to detect the light curve of \dr{} at the ~0\fm05 level
in the $r$ band. 
It would be useful in the future to obtain a further light curve in multiple colours 
($g$ and $r$) to confirm if there is any more subtle colour-dependent variability, 
which could indicate either surface composition or topographic variation.

\section{Characteristics derived from visual range and infrared measurements}

\subsection{Absolute brightness, {colours} and phase correction}

We used the MPG/ESO 2.2m to calculate H$_V$ of \dr{} as well as
66 points of V-band data in the Minor Planet Center database as auxiliary data
to calculate the absolute brightness of \dr{}. As these data points
cover the phase angle range of 1\fdg0$<$\,$\alpha$\,$<$3\fdg7, we
were able to fit the slope parameter with a straight line and we obtained 
$\beta$\,=\,0.137$\pm$0.089. We assigned a general error bar of 0\fm3 
to the MPC V-band data points, the median difference reported between
MPC and well-calibrated photometry \citep{Romanishin+Tegler,Benecchi2011}.

%

The MPG/ESO 2.2\,m telescope observations provided 
{absolute magnitudes and colours, as listed in Table~\ref{table:absmag}.}
Using the V-band brightness, the geometry information at the
observation epoch and the $\beta$ value we obtained from 
MPC data we calculated the V-band absolute 
brightness of \dr. The heliocentric and observer-to-target
distances and the phase angle of the observation were 
r\,=\,14.678\,AU, $\Delta$\,=\,14.70\,AU and $\alpha$\,=\,3\fdg952, 
respectively. Based on the observed brightness of 
V\,=\,19\fm254$\pm$0.023 we obtained H$_V$\,=\,7\fm04$\pm$0\fm35.

\begin{table}
\begin{tabular}{ccc}
\hline
Band	&  Magnitude  &  Uncertainty \\  
\hline
B & 19.901 & 0.030 \\
V & 19.254 & 0.023 \\
R & 18.691 & 0.025 \\
I & 18.269 & 0.026 \\
Z & 18.900 & 0.075 \\
\hline
B--V & 0.647 & 0.038 \\
V--R & 0.563 & 0.034 \\
R--I & 0.422 & 0.036 \\
B--I & 1.632 & 0.040 \\
\hline
\end{tabular}
\caption[]{Absolute magnitudes and colours derived for \dr{} 
from the MPG/ESO~2.2\,m observations.}
\label{table:absmag}
\end{table}


\subsection{NEATM models of the thermal emission}

We used the Near-Earth Asteroid Thermal Model (NEATM, Harris 1998) to 
estimate the main characteristics of the target combining the fluxes 
of its reflected light and thermal infrared emission. 
The input fluxes used for the NEATM model (as well as 
for the thermophysical model discussed in the next subsection), 
are calculated from the observed fluxes in the way described in
Sect.~\ref{sect:fluxunc}. 

In our NEATM modelling the beaming
parameter $\eta$, in addition to the effective diameter and geometric albedo,
was treated as a free parameter and therefore fitted to our data points. 
The quality of the fit is characterised by the reduced $\chi^2$ 
values \citep[described e.g. in][]{Vilenius2012}.

\begin{figure}[ht!]
\includegraphics[width=8.5cm]{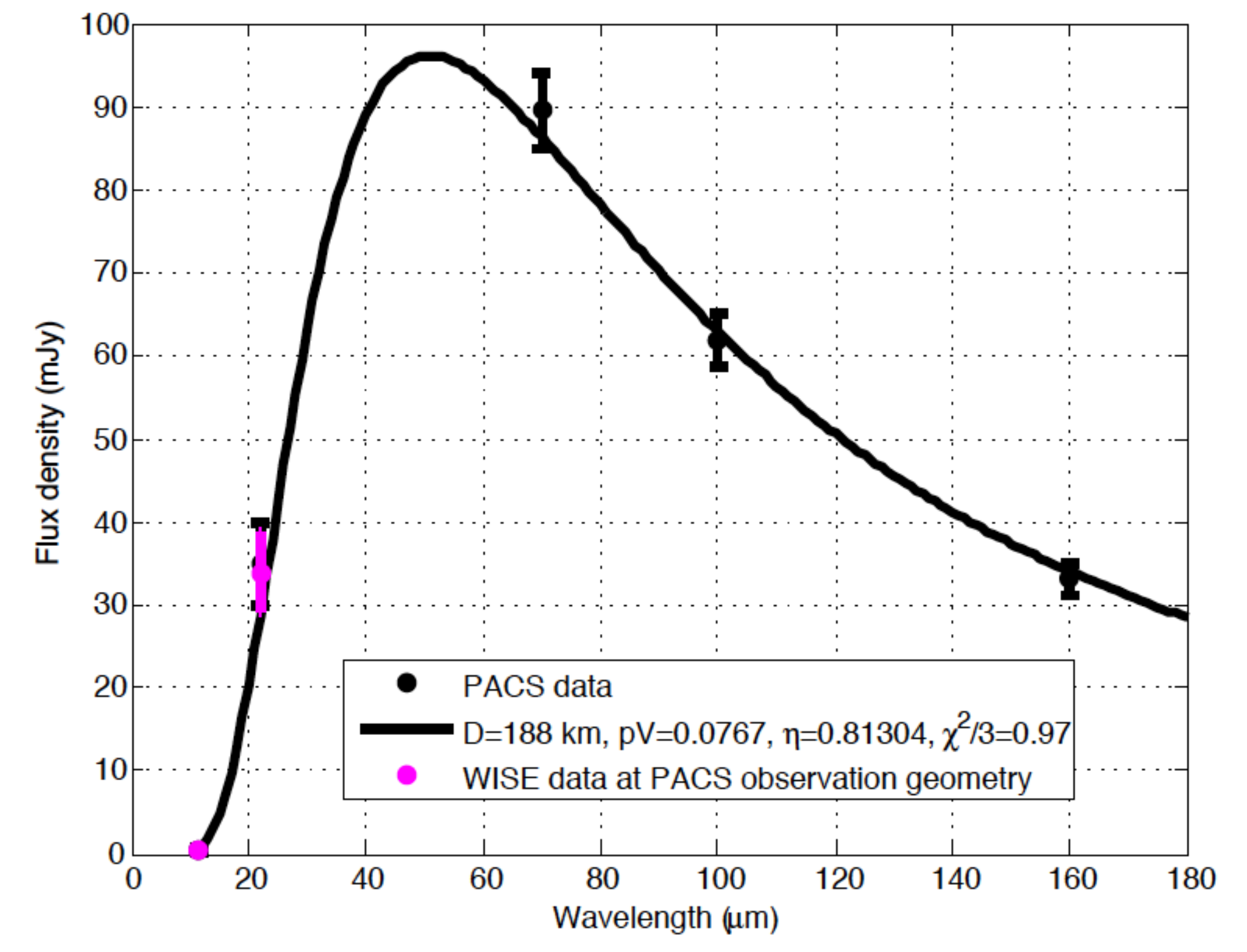}
\caption[]{Best-fit NEATM model considering the Herschel/PACS and
WISE 11\,$\mu$m and 22\,$\mu$m input fluxes. The WISE fluxes 
were transformed to the PACS observations geometry.}
\label{fig:bestfitneatm}
\end{figure}

We considered two sets of data points. In the first one we used the 
combined, "double differential" Herschel/PACS fluxes in the 70, 100 and
160\,$\mu$m bands (three data points, see Table~\ref{table:inputfluxes}), 
while in a second set we used the 11 and 22\,$\mu$m WISE fluxes as well 
(altogether five data points). The "best-fit" result is presented in 
Fig.~\ref{fig:bestfitneatm}. The "PACS only" fit provides 
the best-fit parameters of D$_{\rm eff}$\,=\,173$\pm$17\,km, 
p$_V$\,=\,9.1$_{-2.7}^{+4.4}$\% and $\eta$\,=\,0.57$_{-0.21}^{+0.28}$. 
Using the five "PACS+WISE" data points the best-fit quantities
are D$_{\rm eff}$\,=188.0\,$\pm$9.4\,km, 
p$_V$\,=\,7.6$_{-2.5}^{+3.1}$\% and $\eta$\,=\,0.813$_{-0.062}^{+0.074}$. 

The low value of the $\eta$ parameter ($\sim$0.8) is very close to 
the canonical value 0.756 used for main belt 
asteroids and is different from the
mean value of $\eta$\,=\,1.20$\pm$0.35 in the trans-Neptunian population 
\citep{Stansberry2008,Lellouch2012}. { Trans-Neptunian objects, 
however, show a rather wide range of beaming parameters. 
Recent results indicate an average value of $\eta$\,=\,1.11$\pm$0.15 for 
Plutinos \citep{M2012}, $\eta$\,=\,1.14$\pm$0.15 for 
Scattered Disk Objects \citep{SS2012} and $\eta$\,=\,1.47$\pm$0.43
for Classicals \citep{Vilenius2012}. Although the beaming parameters
derived for Centaurs are rather similar to those of other trans-Neptuinan 
object classes 
\citep[with a median value of $\eta$\,=\,1.12$\pm$0.38][]{Lellouch2012}, 
there are a few Centaurs with $\eta$ values close to or below that of \dr{}, 
down to $\eta$\,$\approx$\,0.4. 
}

\subsection{Thermophysical modeling of the infrared emission}

We also used a thermophysical model (TPM) approach 
\citep[][and references therein]{Muller+Lagerros1998,Muller+Lagerros2002}
to obtain the main surface
characteristics of our target (size, albedo, thermal inertia, surface roughness), 
based on the Herschel/PACS and WISE data
\citep[for details of the present model see][]{MuellerSDP}. 
As the object was bright, and photometry could be performed with 
a relatively high signal-to-noise ratio, we used the thermal fluxes 
of the two epochs independently (the differences in the 
fluxes at the two epochs might reflect rotational variations).  
Unfortunately the rotation period could not be inferred from the
light curve observations (see Sect.~\ref{sect:lightcurve}).
The results we obtained show that the object has a low thermal inertia below 
4\,\ti (assuming P$_{sid}$\,=\,6\,h) or below 9\,\ti (P$_{sid}$\,=\,24\,h), 
except if we have seen it pole-on (in this latter case 
our model is not able to provide any constraints on the object's 
thermal inertia). 
{These two rotation periods encompass the majority 
of the known TNO/Centaur rotation periods. We note from this
that the influence of the rotation period on the derived results 
for \dr{} is very minor}.
A low surface roughness (very smooth surface) is not compatible with the
 observed fluxes, independent of spin-vector orientation, rotation
 period or thermal inertia. The possible size range is 183\,-\,198\,km, 
 using the requirement that reduced $\chi^2$-values can be allowed up to 
 $\chi^2_r$\,=\,1.38 in the case of five independent measurement points. 
 The possible albedo range is 0.060\,-\,0.085, allowing the same 
 $\chi^2_r$ range and also including a $\pm$0\fm1 error for the H-magnitude
 Allowing for the full $\pm$0.35 mag error
 for H$_V$ (see Section 3.1) would lead to a possible albedo
 range of 0.055 -- 0.111 in our TPM analysis. 
The best fit TPM solution
for equator-on geometry provides 
 $\Gamma$\,=\,0.4\,\ti, high roughness, D\,=\,184.1\,km and 
 p$_V$\,=\,0.078, with a corresponding \redchi
value of 0.58.
The pole-on situation produces an even better fit with a
 {\redchi}-value of 0.51, and with the corresponding size and albedo
ranges of D\,=183\,-\,186 km and p$_V$\,=\,0.070\,-\,0.085 while in 
this case the thermal inertia is not constrained.

\begin{figure}[h!]
\vskip -0.8cm
\includegraphics[angle=90,width=8.5cm]{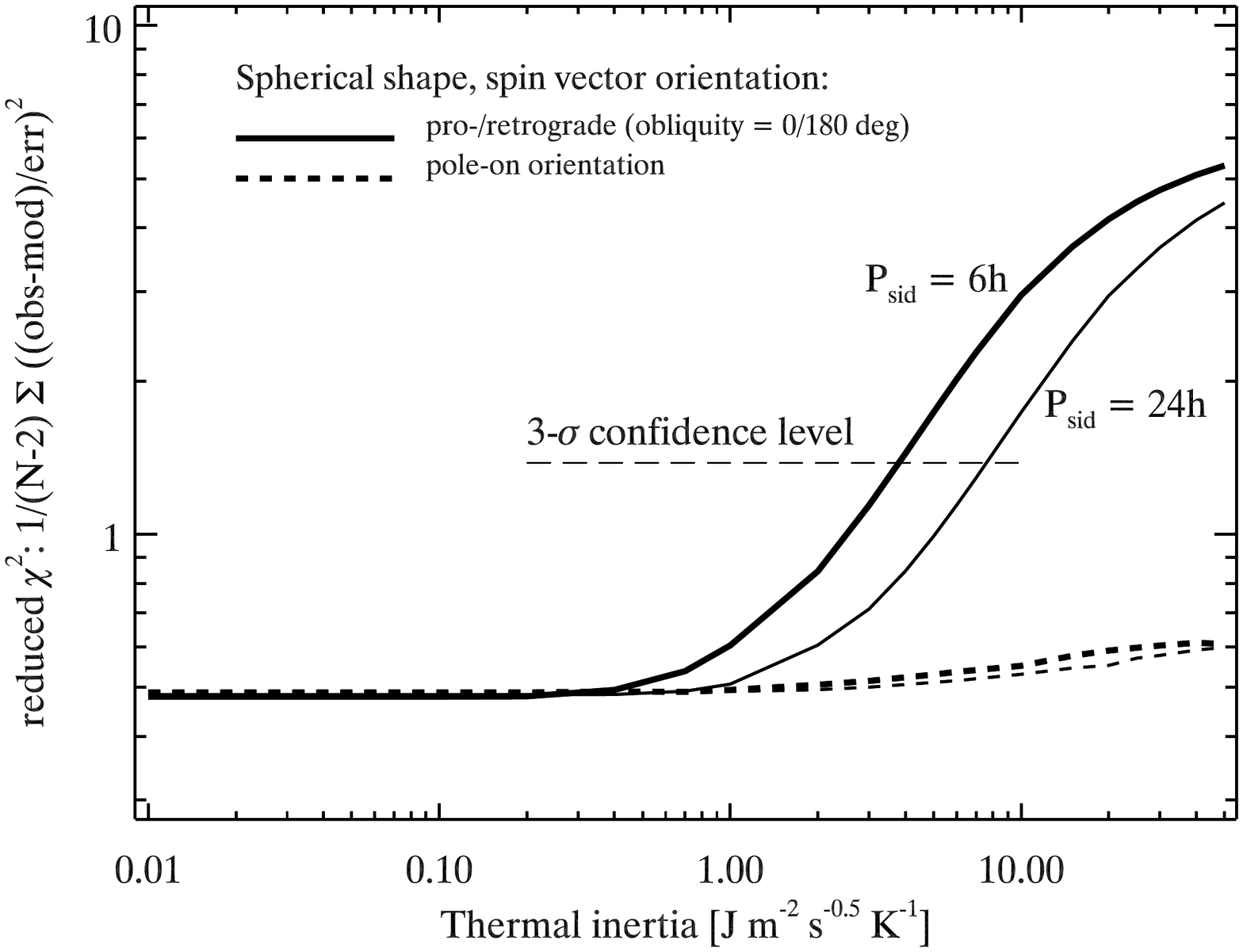}
\vskip -1.3cm
\includegraphics[angle=90,width=8.5cm]{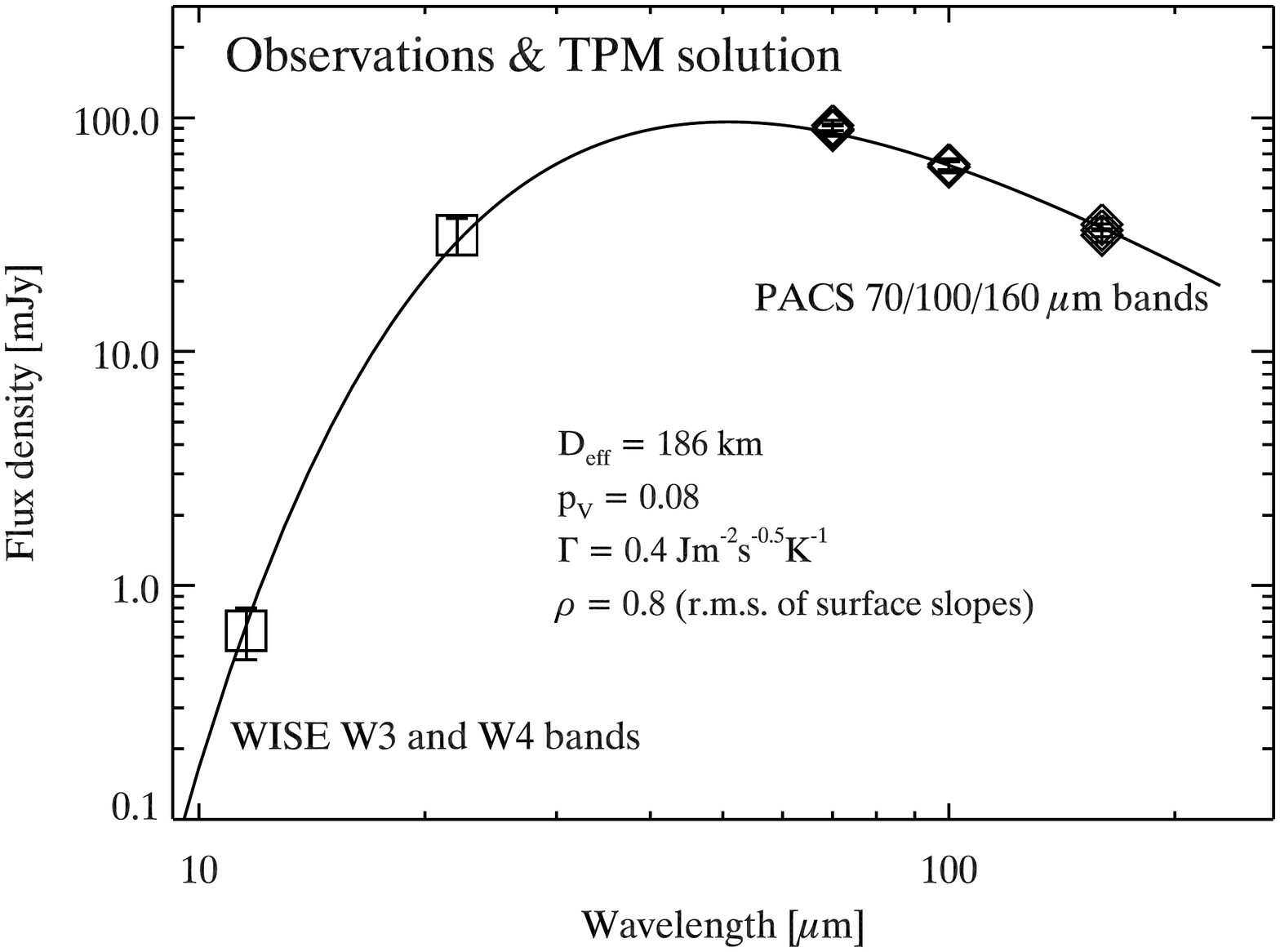}
\vskip -1.3cm
\includegraphics[angle=90,width=8.5cm]{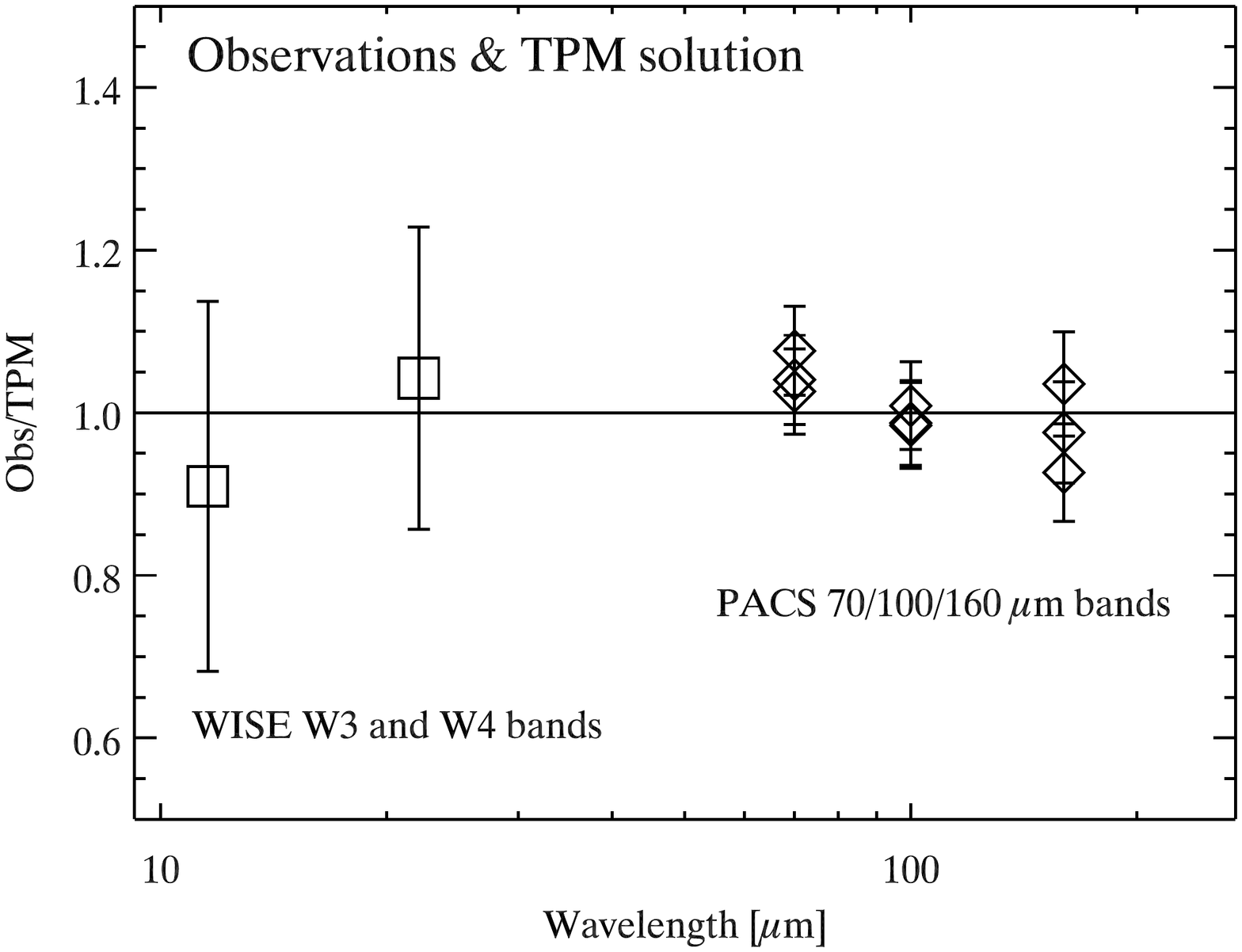}
\vskip -0.5cm
\caption[]{(a) TPM analysis of all available thermal measurements.
 The reduced $\chi^2$-values are shown for two different
 spin vector orientations: equator-on (solid lines) and
 pole-on (dashed lines), each time for two different
 values for the rotation period. The 3-$\sigma$ confidence
 level for the TPM fit to the observations is also shown.
 (b) the best TPM solution is shown together with the observed fluxes
 (c) the same model, but now shown in the observation/TPM picture.}
\end{figure}

\subsection{Colours and visual reflectance \label{sect:refl}}

\begin{figure}
\includegraphics[width=8cm]{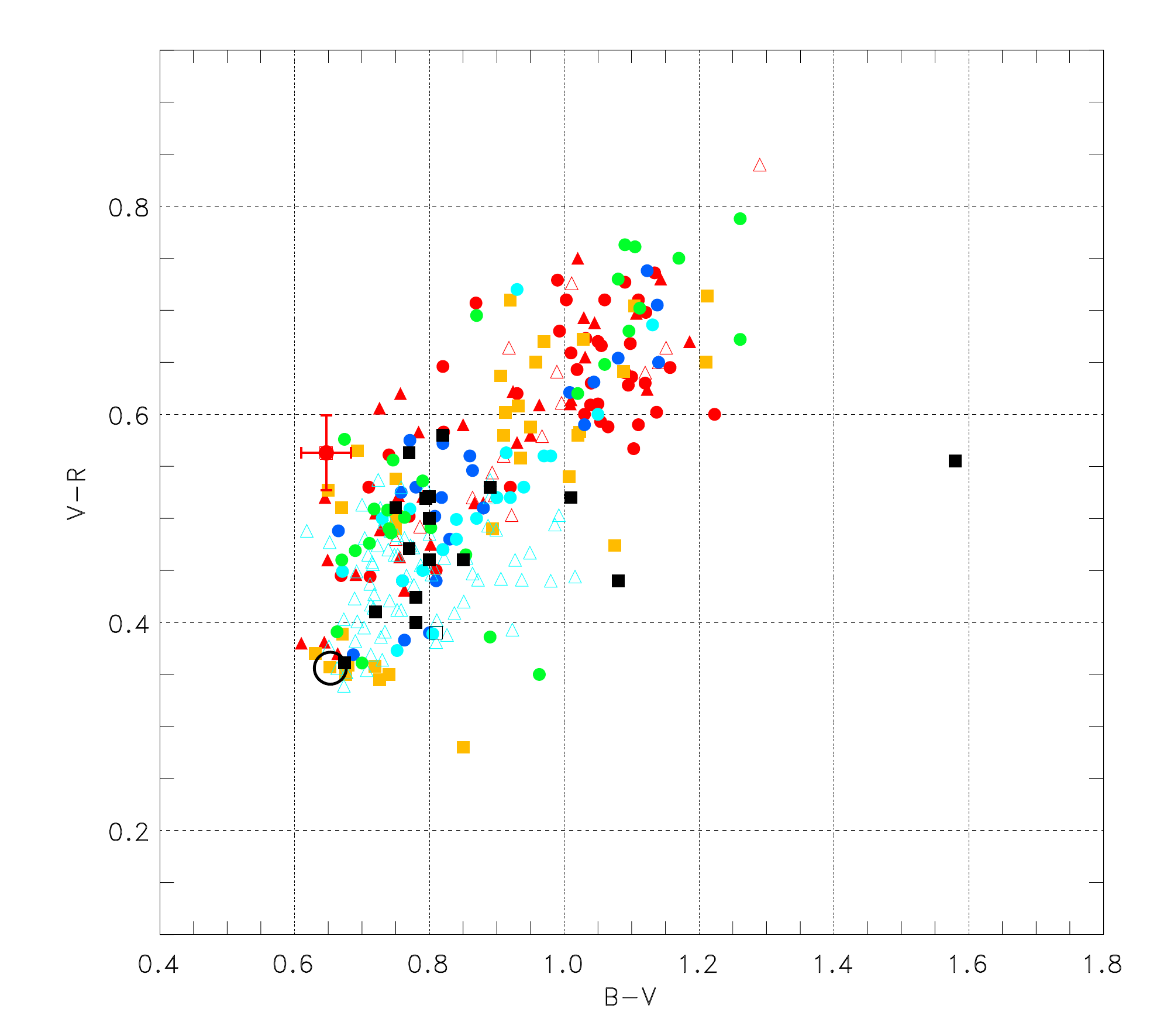}
\includegraphics[width=8cm]{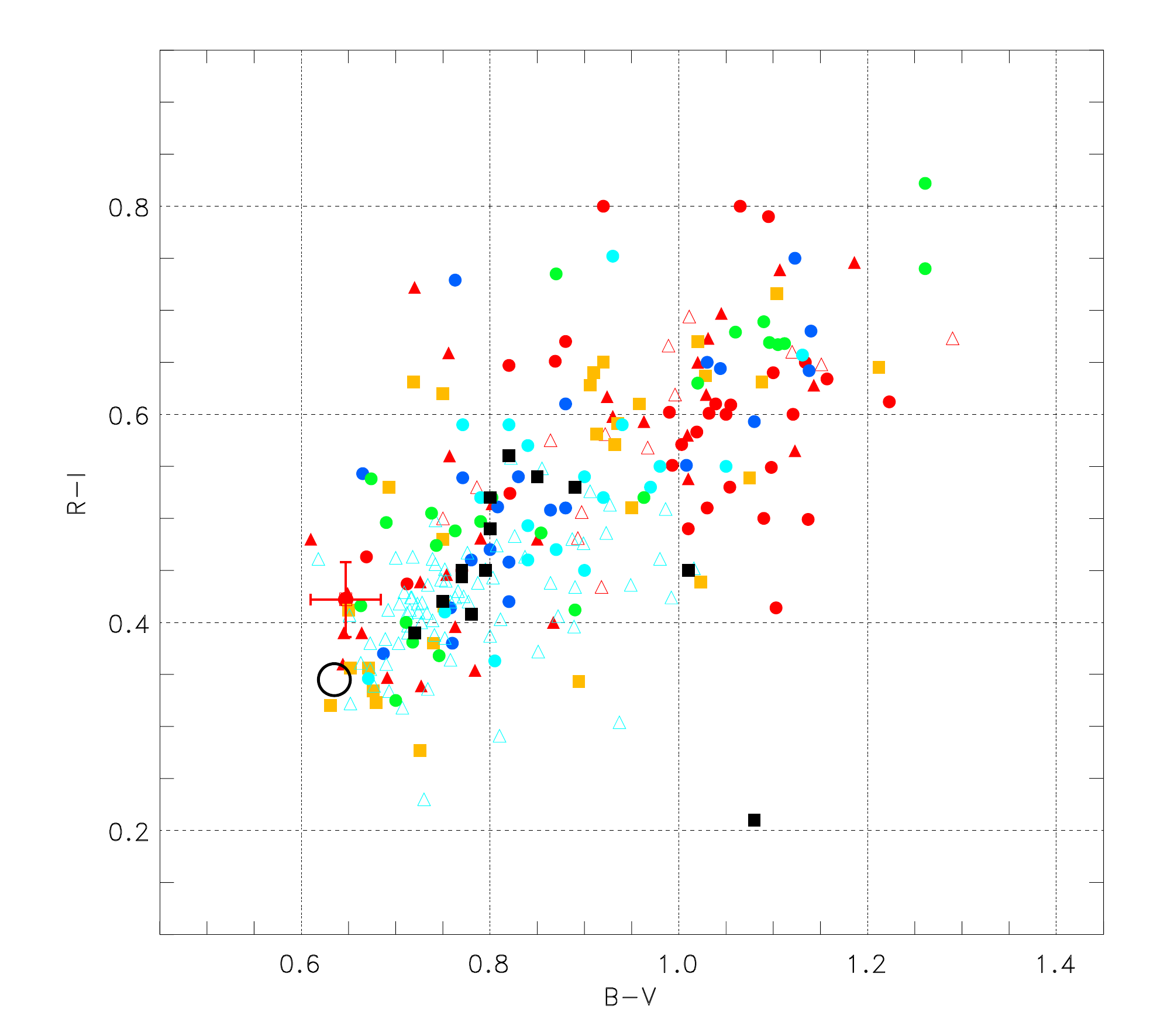}
\caption[]{ {\it Upper panel}: B-V versus V-R colours of \dr{} and those of the objects
in the "MBOSS2" database \citep{mboss}. The symbols correspond to
various dynamical classes of the trans-Neptunian object population and of
comets, as follows:  
(1) red filled triangle: plutions; 
(2) open red triangle: other resonant objects;
(3) filled red circle: cold Classicals; 
(4) filled orange square: hot Classicals;
(5) filled yellow square: other Classicals;
(6) filled dark blue circle: Scattered disk objects;
(7) filled light blue circle: Detached objects;
(8) open light blue triangle: Jupiter Trojans;
(9) filled green circle: Centaurs;
(10) filled black square: short period comets;
(11) open black square: long peroid comets. 
The big black open circle represents the solar colours
\citep{Ramirez2012}. 
{\it Lower panel}: The same as the upper panel, but now presenting the B--V versus
R-I colours of \dr{} and the objects in the "MBOSS2" database.}
\label{fig:mboss}
\end{figure}

{We plotted the B--V vs. V--R and B--V vs. R--I colours of 
\dr{} along with those of other outer Solar System objects
using the MBOSS-2 database \citep[see fig.\ref{fig:mboss} in][]{mboss}. 
Compared to the colours of other bodies, 
\dr{} is certainly "blue" when its B--V colour is considered 
(close to the solar value), while it shows colours closer to 
the population average in R--I, and especially in R--I. 
%
}

{When comparing our observations of \dr{} to the observed 
colours of the Centaurs alone (the green filled circles in 
Fig.~\ref{fig:mboss}), we find that \dr{} has a noticeably 
lower B--V colour than any of the Centaurs in the MBOSS sample. 
In the other colours (V--R and R--I) its colours are not vastly different
to those of the Centaurs, but are close to the "blue" side of the distributions. 
We note that the colours of \dr{} are in general very close to 
those of 2002\,DH$_5$. }

\begin{figure}[ht!]
\includegraphics[width=8.5cm]{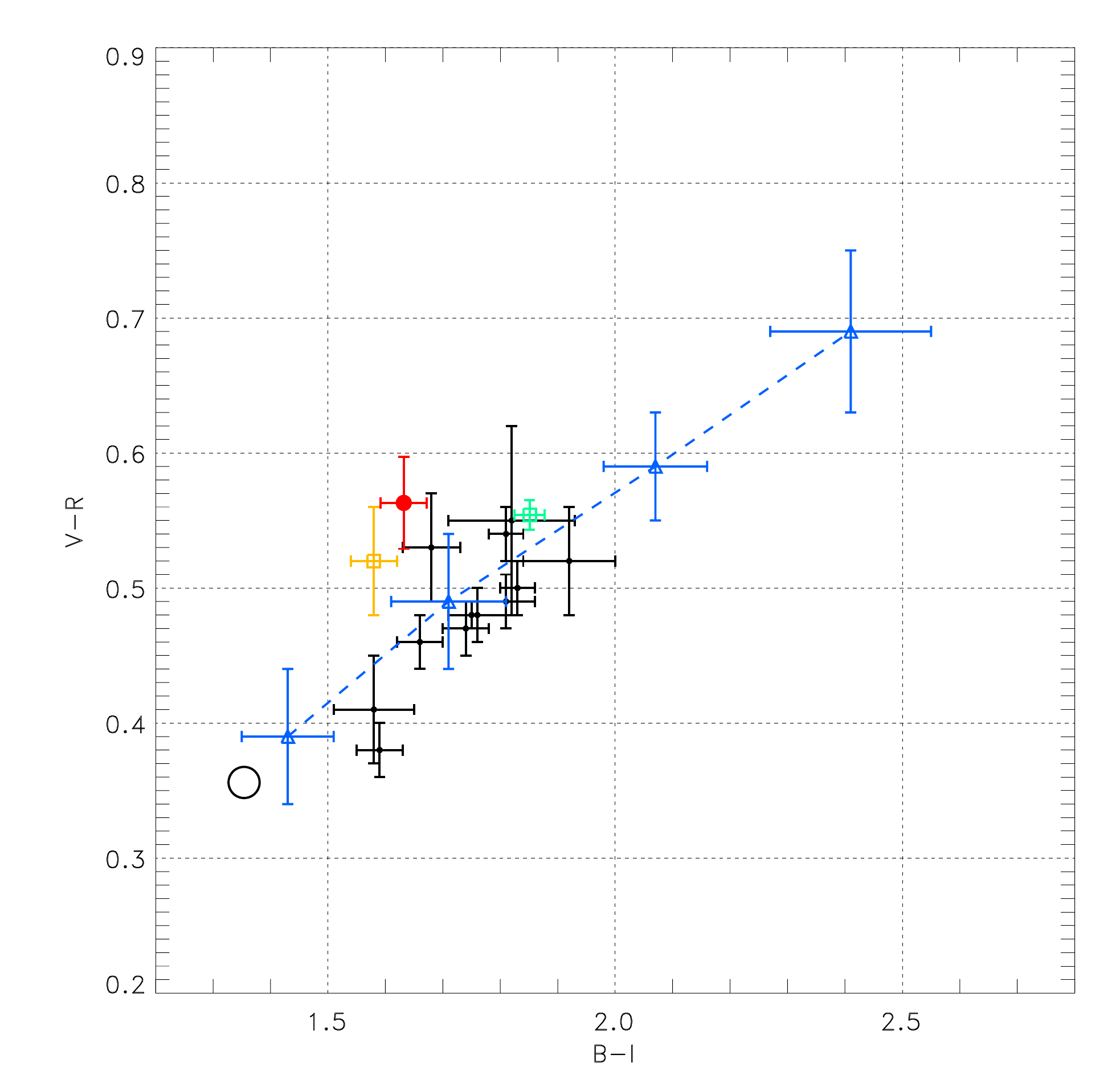}
\caption[]{(B-I) versus (V-R) colours of \dr{} (red filled circle with error
bars) and some representative object types. Black data points represent 
the colours of some Damocloids by \citet{Jewitt2005}; blue triangles
with errors bars correspond to the mean values of the BB, BR, RI and RR
taxonomy classes of TNOs and Centaurs \citep{Perna}, from bottom-left to
top-right, respectively; the green and orange points show the median colours
of the olivine bearing S-type and V-tpye asteroids, respectively
\citep{Chapman1993}. The big black open circle marks the solar colours
\citep{Ramirez2012}. 
{The erros bars of the individual Damocloid points correspond to the
errors of the colour determination \citep{Jewitt2005} while in the 
case of the taxonomy class median values they represent the standard
deviation of the distribution. 
}}
\label{fig:colors}
\end{figure}

As dynamically \dr{} {would also fit} into the group of Damocloids 
(black symbols in Fig.~\ref{fig:colors}) we also compared
the colours of these objects with those of \dr{}.  
While Damocloids  seem to follow the main colour trend 
(blue dashed line) of the TNO taxonomy classes, the 
colours of \dr{} are rather different form those 
of the other group members. In addition we also plotted
the average colours of S- and V-type asteroids. The colours
of \dr{} are definitely different from the S-type colours, 
but are close to the colours of V-type asteroids
(green and orange symbols in Fig.~\ref{fig:colors}, respectively). 
 
\begin{figure}[ht!]
\includegraphics[bb=45 27 484 341,width=8.5cm]{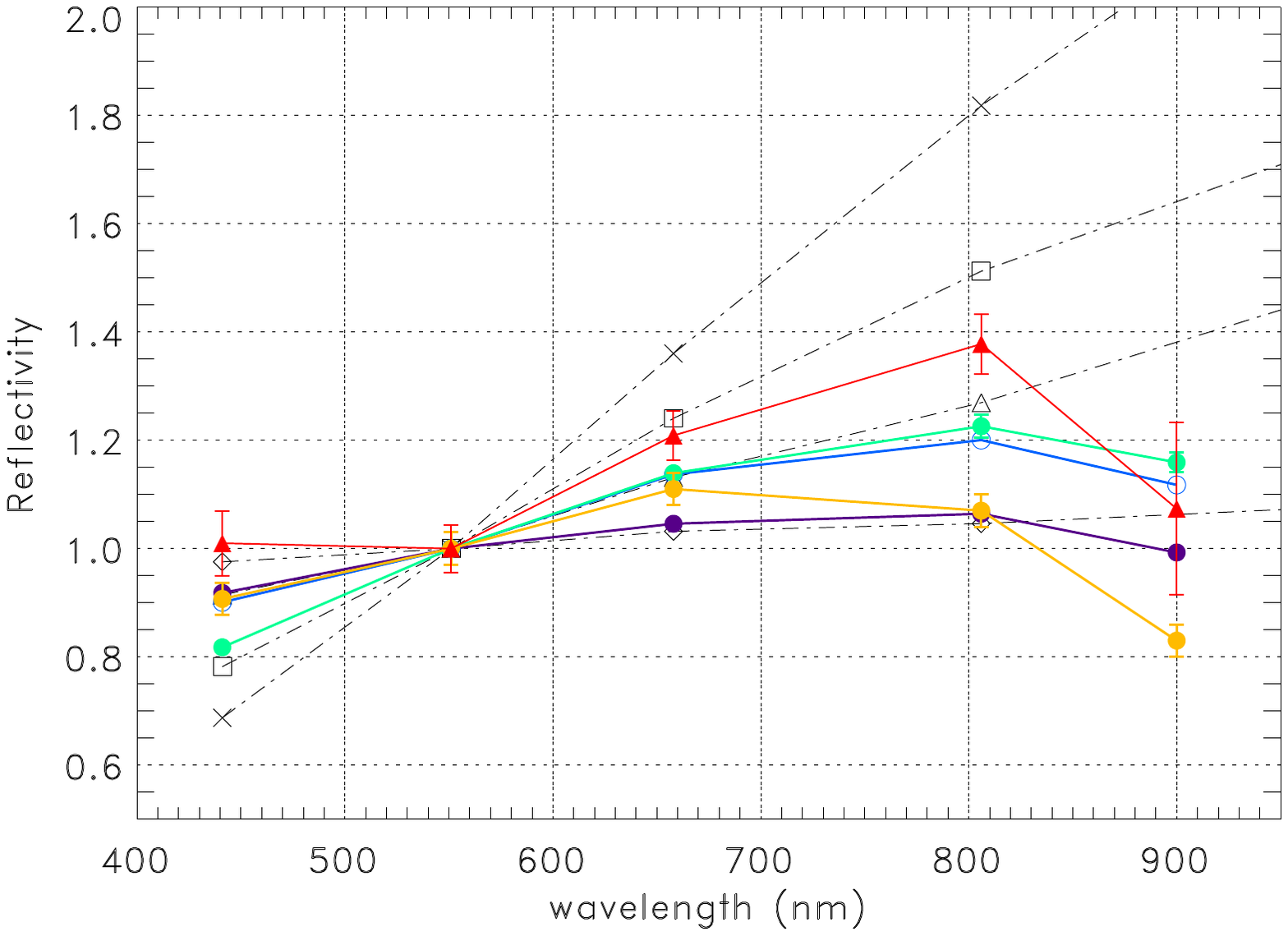}
\smallskip
\caption[]{Normalized spectral reflectivity of \dr{} in the optical 
(red curve), in comparison with the main TNO taxonomy classes BB, BR, RI, and RR 
\citep[dashed-dotted lines from bottom to top,][]{Perna}, and the TNOs Eris 
\citep[purple curve][]{AlvarezCandal} and Makemake \citep[dark blue curve][]{Makemake}. 
In addition, the average reflectivities of the S- and V-type asteroids
are also shown \citep[green and orange curves, respectively,][]{Chapman1993,arefl2}. 
Note the $Z$-band feature of \dr{} is unseen in the usual TNO taxonomy classes 
and could most readily be attributed to the presence of a reddened 
olivine-bearing surface (see the text for details).}
\label{fig:refl}
\end{figure}

Using the absolute brightness values derived from the MPG/ESO 2.2\,m
measurements we calculated normalized reflectance using 
the solar colours of the SDSS magnitude transformation 
page\footnote{http://www.sdss.org/dr5/algorithms/sdssUBVRITransform.html}. 
\dr{} is represented by the red curve in Fig.~\ref{fig:refl}. 
The most obvious feature one can identify is the
presence of a strong Z-band absorption feature which is 
not seen in any of the main TNO taxonomy classes. 
Some objects with methane on their surfaces show absorption
in the Z-band due to the 890\,nm CH$_4$ line. However, these are
large and very high albedo objects, like Eris and Makemake
\citep[][respectively]{AlvarezCandal,Makemake}. 
The normalized reflectivity curves of these objects are plotted
in Fig.~\ref{fig:refl} as well. As \dr{}-s geometric albedo 
is $\sim$8 per cent only, it would be very challenging to construct 
a surface composition which can reproduce the observed reflectivity, since
even a smaller amount of methane could increase the albedo considerably 
and this is incompatible with the present albedo of the object. 

Z-band absorption, however, can be relatively easily reproduced if it is
due to the presence of olivine or pyroxene 
(with the strongest absorbance at $\sim$1\,$\mu$m),
as it is the case in S- and V-type asteroids. Concerning just 
the depth of the $Z$-band absorption,
the reflectance of V-type asteroids ressembles the most to that of \dr{}, 
however, with a notably different spectral slope at the shorter wavelengths. 
On the other hand, reflectivities of V-type asteroids may be modified by space weathering 
effects \citep{Hiroi,Binzel}, 
resulting in a reflectance more similar to that of \dr{}. A-type asteroids 
have similar reflectance spectra due to olivine \citep{DeMeo2009}.




\section{The dynamics of \dr{} \label{sect:dyn}}


In order to examine the dynamical behaviour of \dr{}, we used the Hybrid integrator 
within the n-body dynamics package MERCURY \citep{Chambers} to follow the evolution 
of the orbit under the gravitational influence of Jupiter, Saturn, Uranus 
and Neptune for a period of 4\,Gyr into the future. Following a procedure established 
in earlier works 
\citep[][]{Horner2004,Horner+Lykawka2010b,Horner2012a,Horner2012b}
we created a suite of test particles distributed in even sized steps across the 
3-sigma error ellipses of the object’s best-fit orbit in perihelion distance, 
$q$, eccentricity, $e$, and inclination, $i$. In this way, we created a grid of 
45$\times$45$\times$45\,=\,91,125 test particles in q-e-i space, 
centred on the nominal best-fit orbit for \dr{}. Each of these test particles was 
then followed in our integrations until it was removed from the system, either by 
colliding with one of the massive bodies (i.e. the Sun or one of the giant planets) 
or being ejected to a heliocentric distance of 10,000 AU.

\begin{figure}
\includegraphics[width=8.5cm]{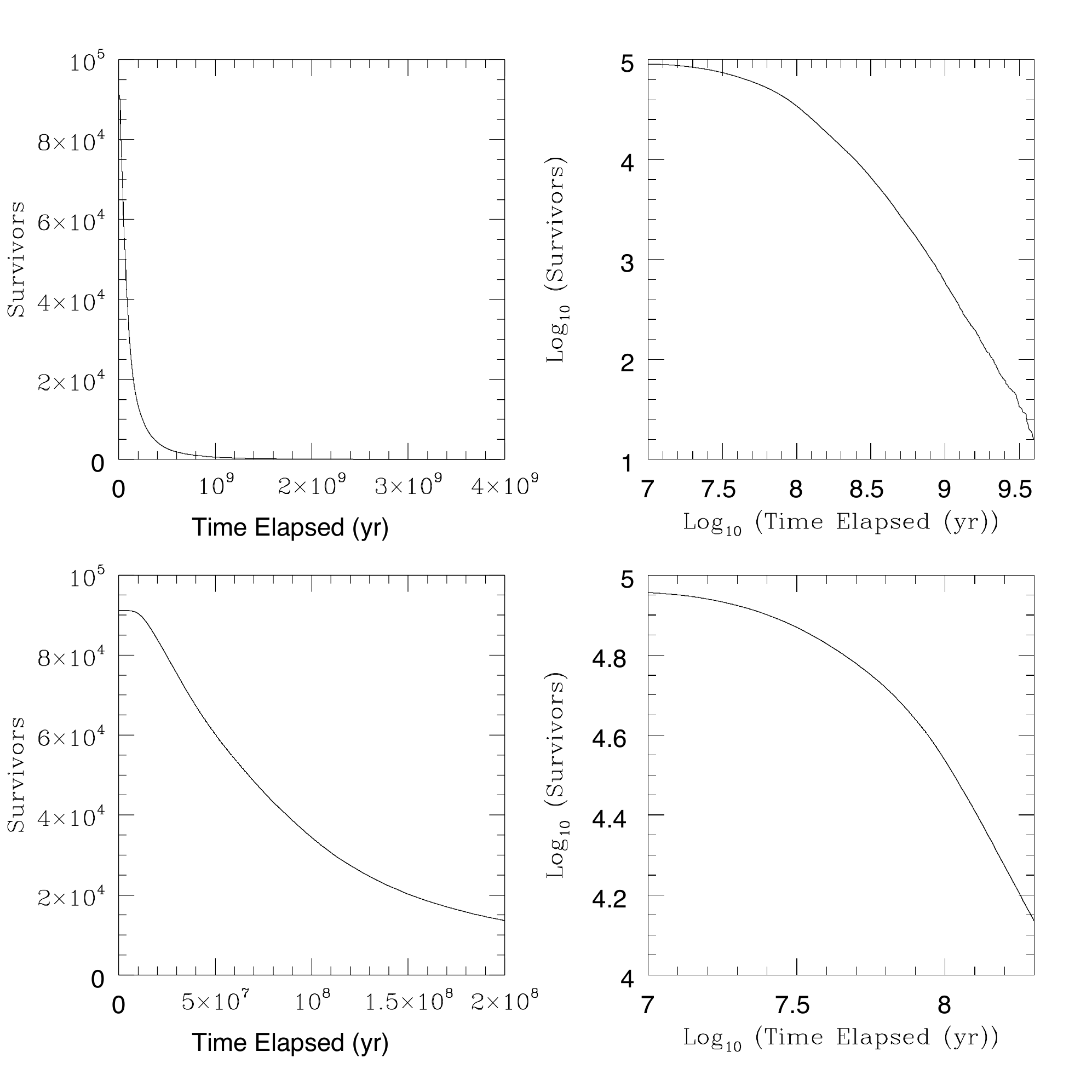}
\caption[]{The decay of our population of 91,125 clones of \dr{} as a function of 
time. The left-hand panels show the decay of the number of surviving clones as a 
function of time, whilst those to the right show the same data as log-log plots. 
The upper panels show the decay across the full 4 Gyr of our integrations, while 
the lower plots show just the first 200 Myr of the evolution of our test particles. 
It is immediately apparent that \dr{} is moving on a  dynamically highly unstable 
orbit, with fully half the test particles being ejected from the Solar System within 
the first 75.5 Myr of the integrations. Such instability is not unexpected, given that 
\dr{}’s orbit crosses those of Uranus and Neptune. }
\label{fig:decay}
\end{figure}

\begin{figure}
\includegraphics[width=8.5cm]{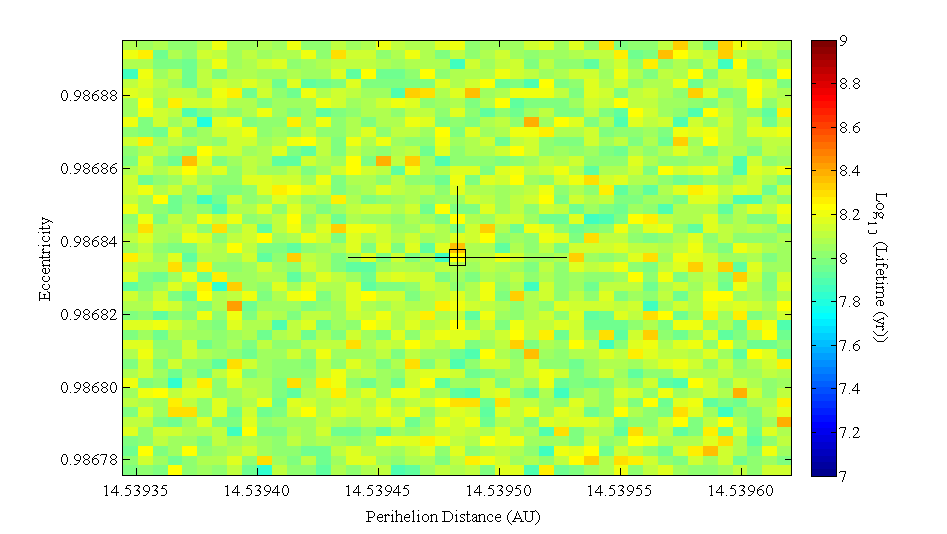}
\includegraphics[width=8.5cm]{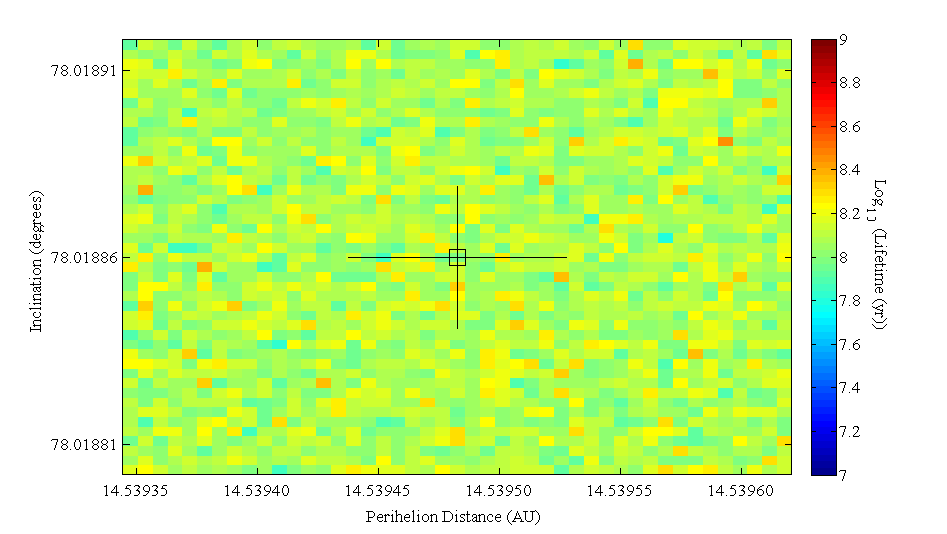}
\caption[]{{\it Upper panel}: The mean lifetime of clones of 2012 DR30, 
as a function of their initial 
perihelion distance and eccentricity. Each square of this plot shows the Log10 
of the mean lifetime of all 45 test particles that began the simulations at 
that particular a--e coordinate. The hollow box shows the location of the nominal 
best-fit orbit for \dr{}, while the black lines that extend from that box show 
the 1-sigma errors in a and e for that best fit orbit. \\
{\it Lower panel}: The same as the upper panel but for inclination 
instead of eccentricity. 
We note that the stability of \dr{} does not vary significantly as a function 
of the initial eccentricity or inclination and perihelion distance used, a reflection 
of the relatively high precision with which the object’s orbit is known.}
\label{fig:stability}
\end{figure}

\begin{figure}
\includegraphics[width=8.5cm]{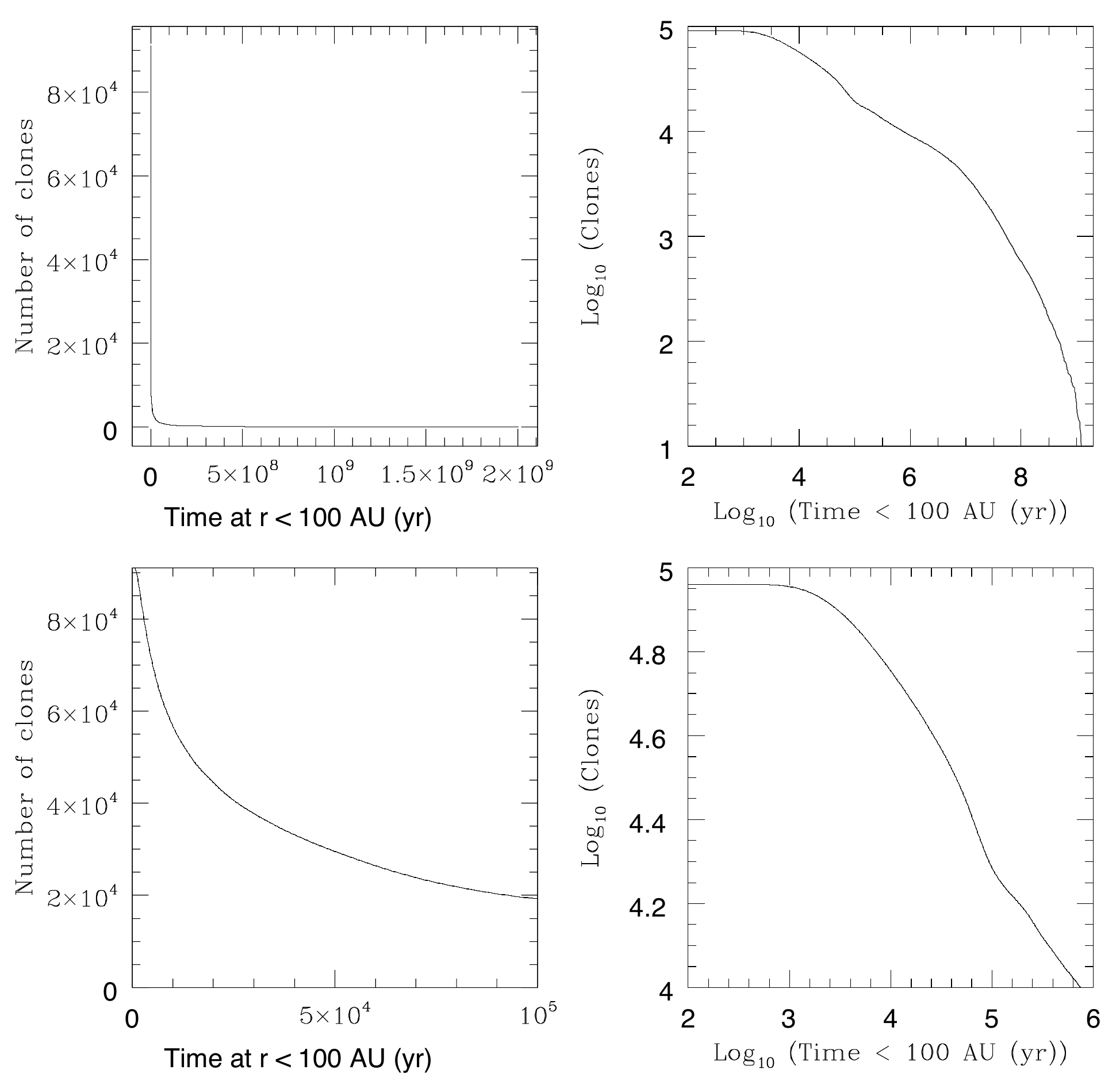}
\caption[]{The total number of clones (Y) that spent at least X years 
closer to the Sun than 100 AU. The left plots show the data 
on a linear scale, while those to the right show it as log-log plots.}
\label{fig:inside100au}
\end{figure}


It is immediately apparent (see Fig.~\ref{fig:decay})  
that \dr{} is moving on a relatively dynamically 
unstable orbit, and that the number of clones that survive, as a function of time, 
decays exponentially. Half of the test particles are removed from the Solar System 
within just 75.5\,Myr. Only 16 of the test particles survived for the full 4\,Gyr of 
our integrations -– just 0.0176 per cent of the total! 
The exhibited instability is independent of the initial perihelion distance, 
eccentricity, and inclination tested (see Fig.~9), although this is not 
hugely surprising, given the remarkably small uncertainties in the orbit of the object. 
Nevertheless, this result is reassuring, in that it tells us that the dynamical 
behaviour we observe for the object is truly representative, in contrast to 
previous studies of Solar System objects such as the Neptun trojan 
2008~LC$_{18}$, whose long-term behaviour was strongly dependent on the initial 
conditions considered \citep{Horner2012a}.

With our large dynamical dataset on the evolution of \dr{}, it is possible to work 
out the frequency with which clones of that object become Earth-crossing objects, 
or how much time they spend approaching the Sun to within a given distance. 
Since dynamical evolution under the influence of gravity alone is a time-reversible 
process, this can then give us some indication of the likelihood that \dr{} has, 
in the past, occupied such orbits, as well as enabling us to estimate how long it 
has already spent within a given heliocentric distance.


Taken over all the clones in our runs, the mean clone lifetime was 124.26\,Myr. 
As the clones of \dr{} evolved, they diffused in orbital element space such that 
10808 became Earth-crossing objects at some point in their evolution, albeit 
typically only for very short periods of time. Just over a third of the test 
particles evolved onto orbits with perihelia under the control of Jupiter 
(i.e. within a heliocentric distance of $\sim$6\,AU), a result entirely in keeping with 
previous studies of the Centaurs \citep{Horner2004}.

Given that \dr{} currently spends the great majority of its time at vast 
heliocentric distances, it is interesting to ask what fraction of its life it has 
likely spent closer to the Sun than a given heliocentric distance. Over the entirety 
of our runs (91,125 test particles with a mean lifetime of 124.26 Myr), we find that 
the mean amount of time clones spent at heliocentric distances of less than 1\,AU was 
just $\sim$1.5 years ($\sim1.2 \times 10^{-6}$\,\% of their lifetime). 
The mean time spent within 10\,AU of the Sun was $\sim$4,400 years 
($\sim3.6 \times 10^{-3}$\,\% of their lives), while the mean 
time spent within 100\,AU was $\sim$3.3 Myr ($\sim$2.7\,\% of their lifetime). 




In terms of the time spent at less than 100 AU, our simulations reveal a wide spread 
of outcomes -– from those objects that spend just a few hundred years within 100\,AU 
before being ejected from the Solar System to those few that spend well over a billion 
years in that region. In Fig.~\ref{fig:inside100au}, 
we plot the number of clones that spend at least a 
certain amount of time within 100\,AU of the Sun, as a function of the total time 
elapsed within that heliocentric distance. 


We note that, although the mean amount of time spent within 100 AU is $\sim$3.3 Myr, 
this mean is heavily biased by a relatively small population of objects that are 
trapped onto relatively long lived orbits within 100\,AU, which contribute vastly to 
the total time spent within that distance. Indeed, the 
\emph{median} amount of time that 
clones spent within 100\,AU of the Sun was just 18.8\,kyr, with 975 of them spending 
under a thousand years within that distance, and almost 21,000 of the test particles
spending under five thousand years within 100\,AU. 

\section{Discussion}

As \dr{} spent a relatively short time at small heliocentric distances 
($<$\,100\,AU) it is interesting to consider whether the surface could have
some memories of the long years spent at the far-out reaches of the Solar System.
Volatiles kept on the surface might be one possibility.

As we mentioned it in Sect.~\ref{sect:refl}, the 890\,nm methane 
absorption band is at the right position to explain the strong Z-band
absorption of \dr{}. However, this possibility is ruled out by
the relatively low albedo of the object. In addition to this,
for an object with the size of \dr{} the methane volatility limit is at a 
$\sim$100\,AU distance from the Sun \citep{Brown2011}. Although \dr{} spends the vast
majority of its lifetime beyond this heliocentric distance, in those
short periods when it is close to its perihelion (closer than $\sim$100\,AU)
the volatile escape rate of methane is so large that \dr{} certainly 
cannot retain this molecule on the surface even for a few thousand years, 
considering either Jeans or hydrodynamic escape rates 
\citep{Schaller,Levi+Podolak}. 
Replenishment of methane from subsurface resources is indeed a
possibility, however, no cometary activity has been observed so far which 
otherwise would support this scenario. 
When mixed with or diluted in other ices (e.g. H$_2$O), 
the escape rate of methane could be significantly different. 
But even the clathrate hydrate of methane has a stability limit of 
53\,K at p\,$\simeq$\,1\,nbar, and hence it cannot survive at the current
distance of \dr{}, about 15\,AU \citep{Gautier2005}. 
In addition, dilution of methane
in other ices would decrease the depth of the absorption features.

Another reason to rule out methane on the surface could be that 
if \dr{} was originated from the Oort cloud and not from the trans-Neptunian 
region, then  cosmic ray impacts on the surfaces so far away from the solar 
magnetosphere protection would be destructive for CH bonds and 
hence methane could not survive.

The very likely lack of methane on the surface favors a scenario in which 
the Z-band absorption is due to e.g. olivine or pyroxene, 
like in V-type asteroids -- if the object were really V-type, 
this would certainly suggest a main belt origin. 
V-type asteroids are usually believed to be originated as impact
ejecta from Vesta itself \citep{Binzel+Xu}, but due to its large size, 
it is very unlikely that \dr{} could be one of them (as indicated by its size, 
\dr{} might be a differentiated object itself). 
Probably the same is true for the relation of \dr{} to the 
very rare A-type asteroids that show similar reflectance spectra with 
strong olivine absorption, and are supposed to come from a completely 
differentiated mantle of an asteroid \citep{DeMeo2009}. 
Note that space weathering may also be an important
factor in shaping the observable spectra and colours of these bodies
\citep{Lucas2012}. 
 
\section{Conclusions}

In this paper we determined the basic physical parameters 
(size, albedo)
of \dr{}, determined its visible colours and also discussed the
dynamics of its orbit. 

Considering dynamical evolution, it  
seems highly unlikely that \dr{} originated within the main belt. 
{The most likely origins are either within the inner Oort cloud 
(as suggested for the high-inclination Centaurs by e.g. Emel'yanenko et al., 
2005 and Brasser et al., 2012) or the outer Oort cloud (following the 
classical cometary capture route put forth by e.g. Wiegert \& Tremaine, 1999).
}
Despite the fact that it is highly unlikely, a main belt origin 
seems to explain more readily our observations that indicate a space-weathered 
V-type asteroidal surface, as discussed in the previous section. 
However, it is a question whether such a surface could be the 
result of a long time exposure of Galactic cosmic rays in the inner Oort cloud, 
beyond the protection of the heliosphere. 
A reflectance spectrum of \dr{} would be highly desirable to
confirm and better characterise the Z-band absorption feature and likely rule out
some of these possiblilties. 

Both the "PACS+WISE" NEATM and the thermophysical model results 
indicated a size of $\sim$185\,km and a V-band geometric 
albedo of 8 per cent for \dr{}. 
With these characteristics, \dr{} is definitely the largest and 
highest albedo Damocloid or high inclination
Centaur ever observed; and it is the fifth largest even among the Centaurs, 
just after 2002\,GZ$_{32}$, Chariklo, Chiron and Bienor 
\citep{Lellouch2012}. 
This size and albedo is rather incompatible with the "extinct Halley-type comet"
picture which is often used to explain the properties of Damocloids. 
The mere existence of \dr{} indicates that objects on Damocloid orbits 
may be of mixed origin and may not just be the once active nuclei of cometary bodies.    

\begin{acknowledgements}
This project was supported by the Hungarian OTKA grants K76816, K83790, 
K104607, the HUMAN MB08C 81013 project of the MAG Zrt., 
the PECS-98073 program of the European Space Agency (ESA) and the 
Hungarian Space Office, the Lend\"ulet 2009/2012 Young Researchers' 
Programs; the Bolyai Research Fellowship of the Hungarian Academy of Sciences, 
and the European Community's Seventh Framework Programme (FP7/2007- 2013) under 
grant agreement no. 269194. 
JLO aknowledges support from grant AYA2011-30106-C02-01 and FEDER funds. 
Part of this work was supported by the German {\it DLR} project 
number~50~OR~1108.

We are grateful to the Herschel Science Centre (ESA/ESOC)
for providing us Director's Discretionary Time for the Herschel Space 
Observatory, as well as to the European Southern 
Observatory through the Max-Planck-Institut f\"ur Astronomie 
for the MPG/ESO 2.2\,m telescope DDT time. 

We are particularly indebted to an anonymous referee for his 
numerous thoughtful comments and suggestions.

\end{acknowledgements}


\appendix
\section{Astrometry and photometry data of the visual range measurements}
\begin{table*}[!ht]
\caption[]{Astrometry of 2012 DR$_{30}$ from Faulkes South 2\,m r'-band 
imaging over four nights in 2012.}
\label{table:dr30astro}
\begin{tabular}{cccccc}
\hline
UT of start of observation & Horizons R.A. & Horizons DEC & observed R.A. & observed DEC & offset\\ 
        & (hh:mm:ss) & (dd:mm:ss) & (hh:mm:ss) & (hh:mm:ss) & (dd:mm:ss) \\
\hline
2012-05-28 08:28:06.355 & 10:16:46.65 & -17:05:44.3 & 10:16:46.47 & -17:05:56.5 & 0:00:12.5 \\
2012-05-28 08:33:26.626 & 10:16:46.65 & -17:05:44.2 & 10:16:46.47 & -17:05:56.4 & 0:00:12.5 \\
2012-05-28 08:38:47.220 & 10:16:46.65 & -17:05:44.2 & 10:16:46.47 & -17:05:56.4 & 0:00:12.4 \\
2012-05-28 08:44:08.066 & 10:16:46.65 & -17:05:44.1 & 10:16:46.47 & -17:05:56.3 & 0:00:12.5 \\
2012-05-28 11:56:14.191 & 10:16:46.65 & -17:05:41.7 & 10:16:46.43 & -17:05:53.8 & 0:00:12.5 \\
2012-05-28 12:01:34.738 & 10:16:46.65 & -17:05:41.6 & 10:16:46.44 & -17:05:53.6 & 0:00:12.4 \\
2012-05-28 12:06:55.562 & 10:16:46.65 & -17:05:41.5 & 10:16:46.44 & -17:05:53.5 & 0:00:12.3 \\
2012-05-28 12:12:16.716 & 10:16:46.65 & -17:05:41.5 & 10:16:46.43 & -17:05:53.5 & 0:00:12.4 \\
2012-05-28 12:17:37.018 & 10:16:46.65 & -17:05:41.4 & 10:16:46.44 & -17:05:53.3 & 0:00:12.3 \\
2012-05-28 12:22:57.712 & 10:16:46.65 & -17:05:41.3 & 10:16:46.44 & -17:05:53.1 & 0:00:12.2 \\
2012-05-30 09:06:03.224 & 10:16:47.13 & -17:05:10.4 & 10:16:46.93 & -17:05:22.6 & 0:00:12.5 \\
2012-05-30 09:11:23.703 & 10:16:47.13 & -17:05:10.3 & 10:16:46.94 & -17:05:22.5 & 0:00:12.5 \\
2012-05-30 09:16:44.081 & 10:16:47.13 & -17:05:10.3 & 10:16:46.94 & -17:05:22.5 & 0:00:12.5 \\
2012-05-30 09:22:04.962 & 10:16:47.13 & -17:05:10.2 & 10:16:46.94 & -17:05:22.4 & 0:00:12.5 \\
2012-05-30 09:27:25.380 & 10:16:47.13 & -17:05:10.1 & 10:16:46.94 & -17:05:22.3 & 0:00:12.5 \\
2012-05-30 09:32:45.671 & 10:16:47.13 & -17:05:10.1 & 10:16:46.94 & -17:05:22.3 & 0:00:12.5 \\
2012-05-30 09:38:06.124 & 10:16:47.14 & -17:05:10.0 & 10:16:46.94 & -17:05:22.2 & 0:00:12.5 \\
2012-05-30 09:43:29.790 & 10:16:47.14 & -17:05:10.0 & 10:16:46.94 & -17:05:22.2 & 0:00:12.5 \\
2012-05-30 09:48:51.063 & 10:16:47.14 & -17:05:09.9 & 10:16:46.95 & -17:05:22.2 & 0:00:12.6 \\
2012-05-31 08:54:21.306 & 10:16:47.75 & -17:04:56.1 & 10:16:47.56 & -17:05:08.2 & 0:00:12.4 \\
2012-05-31 09:10:25.284 & 10:16:47.76 & -17:04:55.9 & 10:16:47.56 & -17:05:08.0 & 0:00:12.5 \\
2012-05-31 09:15:45.913 & 10:16:47.76 & -17:04:55.9 & 10:16:47.56 & -17:05:08.0 & 0:00:12.4 \\
2012-05-31 09:37:09.023 & 10:16:47.77 & -17:04:55.7 & 10:16:47.57 & -17:05:07.8 & 0:00:12.5 \\
2012-05-31 09:42:30.113 & 10:16:47.77 & -17:04:55.6 & 10:16:47.57 & -17:05:07.9 & 0:00:12.6 \\
2012-05-31 09:47:50.256 & 10:16:47.78 & -17:04:55.6 & 10:16:47.58 & -17:05:07.7 & 0:00:12.5 \\
2012-05-31 09:53:11.190 & 10:16:47.78 & -17:04:55.5 & 10:16:47.58 & -17:05:07.7 & 0:00:12.5 \\
2012-05-31 10:14:33.638 & 10:16:47.79 & -17:04:55.3 & 10:16:47.60 & -17:05:07.5 & 0:00:12.5 \\
2012-05-31 10:19:54.273 & 10:16:47.79 & -17:04:55.3 & 10:16:47.59 & -17:05:07.4 & 0:00:12.5 \\
2012-05-31 10:25:15.577 & 10:16:47.80 & -17:04:55.2 & 10:16:47.59 & -17:05:07.3 & 0:00:12.5 \\
2012-05-31 10:30:35.888 & 10:16:47.80 & -17:04:55.2 & 10:16:47.60 & -17:05:07.2 & 0:00:12.4 \\
2012-05-31 10:35:56.461 & 10:16:47.80 & -17:04:55.1 & 10:16:47.59 & -17:05:07.3 & 0:00:12.6 \\
2012-05-31 10:41:17.404 & 10:16:47.80 & -17:04:55.1 & 10:16:47.60 & -17:05:07.1 & 0:00:12.4 \\
2012-05-31 10:46:38.478 & 10:16:47.81 & -17:04:55.0 & 10:16:47.60 & -17:05:07.2 & 0:00:12.5 \\
2012-06-14 10:17:19.847 & 10:17:23.22 & -17:04:27.9 & 10:17:23.03 & -17:04:39.7 & 0:00:12.1 \\
\hline
\end{tabular}
\end{table*}
\begin{table*}
\caption[]{Comparison stars selected in photometric reduction of Faulkes South 
2\,m r'-band imagery of 2012 DR$_{30}$.}
\label{table:compstars} 
\begin{tabular}{lcccc}
\hline
Star id & flux & APASS $m_{r}$ & RA (hh:mm:sd) & Dec (dd:mm:ss) \\ 
\hline
41 & $ 97462.2 \pm 212.4 $ & $16.354 \pm 0.047$ & 10:16:38.30 & -17:10:04.0  \\
61 & $ 113369.1 \pm 219.5 $ & 16.214 & 10:16:56.87 & -17:09:14.7  \\
79 & $ 66749.4 \pm 197.9 $ & & 10:16:57.61 & -17:08:23.0  \\
82 & $ 95622.8 \pm 211.5 $ & & 10:16:33.54 & -17:08:20.4  \\
148 & $ 165926.6 \pm 241.4 $ & $15.629 \pm 0.028$ & 10:16:37.92 & -17:03:09.4  \\
175 & $ 65117.2 \pm 196.9 $ & & 10:16:52.14 & -17:04:19.7  \\
210 & $ 101109.8 \pm 214.0 $ & & 10:16:31.96 & -17:03:57.4  \\
216 & $ 64865.0 \pm 197.0 $ & & 10:16:39.52 & -17:04:55.7  \\
252 & $ 106017.4 \pm 216.1 $ & & 10:16:59.19 & -17:05:36.1  \\
260 & $ 133196.7 \pm 227.9 $ & $15.983 \pm 0.102$ & 10:16:33.02 & -17:06:06.9  \\
273 & $ 66359.9 \pm 197.7 $ & & 10:16:33.29 & -17:06:25.8  \\
\hline
\end{tabular}
\end{table*}

\section{Input fluxes for thermal modelling \label{sect:fluxunc}}

We calculate the colour corrected flux from the measured flux using the C$_{\lambda}$
colour correction factors and the r$_{corr,\lambda}$ flux correction factors. 
The colour correction factors of the Herschel data are
calculated using the actual spectral energy distribution of the
target according to \citet{CC}, while in the case of WISE data
the correction factors are taken from \citet{Cutri}. Then the 
corrected flux is: 
\begin{equation}
F_{\lambda,cc} = r_{corr} {\cdot} F_{\lambda} {\cdot} C_{\lambda}^{-1}
\end{equation} 
and the uncertainty of the corrected flux is:
\begin{equation}
{{{\delta}F_{\lambda,cc}}\over{F_{\lambda,cc}}}  = 
  \, \sqrt{ \bigg( {{\delta F_{\lambda}}\over{F_{\lambda}}} \bigg)^2 + 
  \bigg( {{\delta C_{\lambda}}\over{C_{\lambda}}} \bigg)^2 } 
\end{equation}
where $\delta$C$_\lambda$ is the uncertainty of the colour correction factor.
The final "input flux", used for the modeling of the thermal emission is 
the colour corrected flux, F$_{\lambda,i}$\,=\,F$_{\lambda,cc}$. However, 
the uncertainties of the absolute calibration have to be considered in the 
final "input" uncertainties:
\begin{equation}
\delta F_{\lambda,i} = \sqrt { {\delta}F_{\lambda,cc}^2 + 
   (F_{\lambda,cc} \cdot r_{cal,\lambda})^2 }
\end{equation}
where r$_{cal,\lambda}$ is the calibration uncertainty factor that is given as 
a certain fraction of the measured point source flux for all bands of the
WISE and Herschel/PACS instruments. The actual values of all 
the factors mentioned above are summarized in Table~\ref{table:inputfluxes},
also listing the final input fluxes used the thermal emission modeling.


\end{document}